# A Decade of Piezoresponse Force Microscopy: Progress, Challenges and Opportunities

Sergei V. Kalinin,[1,*] Andrei Rar,[2] and Stephen Jesse[1]

[1] Condensed Matter Sciences Division, Oak Ridge National Laboratory

[2] Metals and Ceramics Division, Oak Ridge National Laboratory

## Abstract

Coupling between electrical and mechanical phenomena is a near-universal characteristic of inorganic and biological systems alike, with examples ranging from ferroelectric perovskites to electromotor proteins in cellular membranes. Understanding electromechanical functionality in materials such as ferroelectric nanocrystals, thin films, relaxor ferroelectrics, and biosystems requires probing these properties on the nanometer level of individual grain, domain, or protein fibril. In the last decade, Piezoresponse Force Microscopy (PFM) was established a powerful tool for nanoscale imaging, spectroscopy, and manipulation of ferroelectric and piezoelectric materials. Here, we present principles and recent advances in PFM, including vector and frequency dependent imaging of piezoelectric materials, briefly review applications for ferroelectric materials, discuss prospects for electromechanical imaging of local crystallographic and molecular orientations and disorder, and summarize future challenges and opportunities for PFM emerging in the second decade since its invention.

---

\* Corresponding author, sergei2@ornl.gov

# I. Introduction

Coupling between electrical and mechanical phenomena is a near-universal feature of inorganic, organic, and biological systems. The simplest example of linear electromechnical coupling is piezoelectricity, in which application of stress results in the electrical polarization (direct piezoelectric effect), while application of electric field results in mechanical displacement (converse piezoelectric effect). Since the discovery of piezoelectricity in the end of the 19th century, piezoelectricity in inorganic materials has been studied in great details, an achievement that was made possible by the combination of macroscopic measurements that provided information on properties and diffraction techniques that elucidate atomic structure, with advanced theory (Fig. 1).[1] From symmetry considerations, piezoelectricity can exist only in non-symmetric polar materials. Another elementary example of electromechanical coupling is electrostriction, in which deformation is quadratic in electric field. Electrostriction is present in all materials; however, in most cases the magnitude of electrostrictive coupling is relatively weak. In piezoelectric and electrostrictive materials, the directionality of the mechanical response is fixed with respect to the lattice.

A more complex example of electromechanically active materials is ferroelectrics, in which polarization and hence directionality of electromechanical activity can be switched by external electric (ferroelectric) or mechanical (ferroelastic) stimuli. The first ferroelectric material discovered approximately 85 years ago was Rochelle salt.[2] In the early forties, the search for materials with high dielectric constants as a substitute for natural mica in capacitor applications led to the discovery of ferroelectricity in the perovskite $BaTiO_3$ simultaneously in the USA, Russia and Japan. Immediately $BaTiO_3$ and related ferroelectric perovskites were recognized as promising materials for the submarine sonar arrays, heralding the beginning of intensive research in the field.[1,3] After the discovery of the piezoelectricity in ferroelectrics, numerous applications as sensors, actuators, transducers, etc. has emerged.[4,5] In the last decade, the developments of deposition techniques for epitaxial ferroelectric thin films and advanced ceramic fabrication have resulted in numerous novel applications such as those in micro- and nanoelectromechanical systems (MEMS).[6,7,8] The ability of ferroelectric materials to exist in two or more polarized states, conserve polarization for a finite period, and change the polarization in a field allows their consideration for non-volatile computer memory devices (FeRAM).[9,10,11]

Extending beyond the realm of inorganic materials, electromechanical coupling is a nearly universal feature of biological systems, examples ranging from piezoelectricity of calcified and connective tissues to voltage controlled muscular contractions,[12] cell electromotility,[13] electromotor proteins,[14] etc. In fact, first observations of electromechanical coupling in biological systems by Galvani performed more than 200 years ago[15] (muscular contraction in a frog under an electric bias) were among the founding experiments in discovery of electricity. Similarly to inorganic materials, the simplest manifestation of the electromechanical behavior in biosystems is piezoelectricity, which stems from the crystal structure of most biopolymers including cellulose, collagen, keratin, etc. Piezoelectric behavior has been observed in a variety of biological systems including bones,[16,17,18,19] teeth,[20] wood,[21,22] and seashells.[23] It has been postulated that the piezoelectric coupling, via mechanical stress that generates the electric potential, controls the mechanisms of local tissue development.[24,25] However, complex hierarchical structure of these materials renders quantitative piezoelectric measurements impossible and even the symmetry of the piezoelectric constant tensor in bones, etc.m has not been unambiguously determined.

To summarize, the tremendous progress achieved in understanding of piezo- and ferroelectricity in inorganic materials in the last 50 years has largely been due to the availability of macroscopic single crystalline samples, for which macroscopic property measurements could be correlated with atomic structure and phonon spectra determined by scattering techniques. At the same time, understanding of electromechanical coupling in nanocrystalline ferroelectrics, thin films, piezoelectric biological materials, and phase separated relaxor ferroelectrics, requires local measurements of piezoelectric coupling on the submicron and nanometer scales. In this review, we summarize some of the prospects for nanoscale electromechanical measurements, discuss recent achievements and challenges in the interpretation of Piezoresponse Force Microscopy (PFM), and discuss future strategies for the development of the field.

### I.1. Piezoelectricity and Chemical Bond

Piezoelectricity refers to a linear coupling between the electrical and mechanical phenomena. The direct piezoelectric effect, $d$, relates polarization, $P$, to stress, $P = dX$, while reverse effect relates strain to electric field, $x = dE$. For linear crystalline piezoelectric, the relationship between strain, displacement, and field are summarized in Table 1.

Table 1. Electromechanical coupling in crystals and molecules

|  | Crystal | Single bond |
|---|---|---|
| Displacement | $x_i = s_{ij} X_j + d_{ik} E_k$ | $x = (1/k) F/l + (2q/kl) E$ |
| Charge | $D_i = d_{ij} X_j + \varepsilon_{ik} E_k$ | $P = (2q/k) F + (2q^2/k) E$ |

From the thermodynamic Maxwell relations, the piezoelectric constants for direct and reverse effects are equal; thus studies of e.g. electromechanical response can provide insight into the polarizability of material and vice verse. Piezoelectricity exists for most non-centrosymmetric polar materials and is thus widely spread in nature. The typical order of magnitude ranges from 1 pm/V for quartz to 1000 pm/V for some ferroelectric materials.

The atomic origins of piezoelectricity are directly related to the bond dipoles.[26] This phenomenon is illustrated in Fig. 2 a,b, showing a polar diatomic molecule under the simultaneous action of external force, $F$, and electric field, $E$. The effective spring constant of the bond is $k$. An electric field acting along the molecule axis induces an electrostatic force, $F_{el} = 2qE$, and the elongation of the molecule is $dl = (F + F_{el})/k = F/k + 2qE/k$. In terms of strain, $x = dl/l$, where $l$ is equilibrium bond length, this relationship can be written as shown in Table 1. Similarly, the polarization generated in response to strain can be derived. Note that similarly to crystalline solid, the piezoelectric constants for direct and reverse effect for single bond are equal.

From this analysis, the piezoelectric constant for a single chemical bond is directly related to the bond parameters as $d = 2q/kl$. Estimating $q = 0.3e$, $k = 100$ N/m, $l = 1$ A, the strength of molecular piezoelectric coupling is $d = 9.6$ pm/V. Note that the magnitude of the piezoelectric effect for a single bond is high and is strongly related to the bond parameters (spring constant, bond length, and bond dipole). For macroscopic systems, interaction between individual bond dipoles can give rise to a broad set of collective phenomena ranging from ferroelectricity in perovskites and certain polymers to flexoelectricity in cellular membranes.[1,27] However, for most piezoelectric materials these collective interactions are relatively weak and piezoelectric properties can be understood using a "charged ball and spring" model similar to that illustrated in Fig. 2.

## I.2. Piezoelectricity as a Probe of Molecular Orientation and Ordering

In addition to piezoelectric and ferroelectric domain imaging in crystalline materials, quantitative local electromechanical measurements open at least two novel venues for characterization of materials nanostructures. Piezoelectricity is described by a rank 3 tensor, and is thus strongly orientation dependent. Thus, quantitative electromechanical measurements can provide information on local crystallographic orientation, i.e. relationship between the coordinate system linked to crystal and laboratory. The coordinate transformation between the two requires three rotations described by the Euler angles $\phi$, $\theta$, and $\psi$. For crystalline materials, the relationship between the piezoelectric constant tensor in the laboratory coordinate system, $d_{ij}$, and the tensor in the crystal coordinate system, $d_{ij}^0$, is:

$$d_{ij} = A_{ik} d_{kl}^0 N_{lj} \qquad , \qquad (1)$$

where the matrices $N_{ij}$ and $A_{ij}$ are functions of the Euler angles.[28] As an example, we consider tetragonal BaTiO$_3$. In the coordinate system of the crystal, the $d_{ij}^0$ tensor is

$$d_{ij}^0 = \begin{pmatrix} 0 & 0 & 0 & 0 & d_{15}^0 & 0 \\ 0 & 0 & 0 & d_{15}^0 & 0 & 0 \\ d_{31}^0 & d_{31}^0 & d_{33}^0 & 0 & 0 & 0 \end{pmatrix} \qquad (2)$$

For a general orientation of the crystal, the response components relevant to PFM are:

$$d_{33} = \left( d_{15}^0 + d_{31}^0 \right) \sin^2 \theta \cos \theta + d_{33}^0 \cos^3 \theta \qquad (3)$$

$$d_{34} = -\left( d_{31}^0 - d_{33}^0 + \left( d_{15}^0 + d_{31}^0 - d_{33}^0 \right) \cos 2\theta \right) \cos \psi \sin \theta \qquad (4)$$

$$d_{35} = -\left(d_{31}^0 - d_{33}^0 + \left(d_{15}^0 + d_{31}^0 - d_{33}^0\right)\cos 2\theta\right)\sin \psi \sin \theta \qquad (5)$$

For materials with rotational symmetry for which response is independent of $\phi$, solutions to Eqs. (3-5) can be represented as piezoresponse surfaces, as illustrated in Fig. 3. The shapes of these response surfaces are strongly dependent on materials symmetry, as illustrated for BaTiO$_3$, PbTiO$_3$, and collagen. For low symmetry materials, elements of $d_{ij}$ depend on all three Euler angles, and surfaces become extremely complex and can be represented only in 4D. However, this strong orientation dependence of electromechanical response provides an approach for mapping local crystallographic orientation, i.e. if the elements of the piezoelectric constant tensor can be experimentally measured, local crystallographic orientation, $(\phi_i, \theta_i, \psi_i)$, can be completely or partially derived.

Piezoelectric coupling can emerge even in the *partially* ordered polar materials, including poled ferroelectric ceramics, ferroelectric polymers, and many biological systems, such as connective and calcified tissues and wood. For such materials, piezoelectric coupling is directly related to degree of ordering. For example, for a texture of disordered tetragonal crystal with axial disorder, the effective piezoelectric coefficients for texture, $d_{ij}^0$, is related the piezoelectric coefficients for the original material, $d_{ij}^1$, as[29]

$$d_{15}^0 = \frac{1}{4}\left(1 + \cos\theta_c\right)\left[\left(1 + \cos^2\theta_c\right)d_{15}^1 + \sin^2\theta_c\left(d_{33}^1 - d_{31}^1\right)\right] \qquad (6)$$

$$d_{31}^0 = \frac{1}{8}\left(1 + \cos\theta_c\right)\left[\left(4 - \sin^2\theta_c\right)d_{31}^1 + \sin^2\theta_c\left(d_{33}^1 - d_{15}^1\right)\right] \qquad (7)$$

$$d_{33}^0 = \frac{1}{4}\left(1 + \cos\theta_c\right)\left[\left(1 + \cos^2\theta_c\right)d_{33}^1 + \sin^2\theta_c\left(d_{15}^1 + d_{31}^1\right)\right] \qquad (8)$$

where $\theta_c$ is a parameter (angular distribution of crystallites) that describes the degree of disorder in texture (Fig. 2 b). The evolution of electromechanical response with degree of disorder in the texture of $BaTiO_3$ crystallites is shown in Fig. 4. Note that with the increasing degree of disorder the shape of the response surface simplifies, becoming similar to that of $PbTiO_3$. For large disorder, the response decreases rapidly, becoming zero for isotropic system, $\theta_c = \pi$. Thus, measurement of electromechanical coupling thus yields a degree of ordering in material. Interestingly, early motivation of studies of piezoelectricity of wood in 1950s was quality control, when the presence of internal defects decreased the piezoelectricity.[22]

## II. Nanoscale Probing of Electromechanical coupling

The development of ferroelectric based nonvolatile computer memory technology, as well as emerging applications described above, necessitates imaging ferroelectric materials and local electromechanical measurements on the nanometer scale. The answer to this challenge has come from the field of Scanning Probe Microscopy (SPM), namely Piezoresponse Force Microscopy. In PFM, a local oscillatory electric field is generated by applying an ac voltage to a conducting tip in contact with a sample, and the deformation due to the piezoelectric effect is detected. The imaging paradigm in PFM is complementary to conventional scanning force microscopies (SFM) and scanning tunneling microscopy (STM): while SFMs are sensitive to tip-surface forces through the mechanical motion of the cantilever (mechanical detection) [30] and STM is sensitive to tip-bias induced current (current detection),[31] PFM detects bias-induced surface displacement (electromechanical detection). In less than a decade since its invention, PFM was established as a powerful tool for probing local electromechanical activity on the nanometer scale.[32,33,34,35] Developed originally for imaging domain structures in ferroelectric materials, PFM was later extended to local hysteresis loop spectroscopy[36,37] and ferroelectric domain patterning for applications such as high density data storage,[38,39] and ferroelectric lithography.[40,41,42] Broad applicability of PFM to materials such as ferroelectric perovskites, piezoelectric III-V nitrides,[43] and recently, biological systems such as calcified and connective tissues[44,45,46] has resulted in constantly increasing number of publications, as illustrated in Fig. 5.

The initial applications of PFM were invariably based on qualitative imaging of domain structures, where morphological information was sufficient for materials characterization. However, PFM spectroscopy, high-resolution imaging, and imaging of piezoelectric materials have brought about the challenge of quantitative electromechanical measurements, necessitating an understanding of relationship between PFM signal and local piezoelectric and elastic constants of material. The primary factors determining imaging mechanism in PFM are:

**1. Voltage dependent contact mechanics of tip surface junction.**

Local electromechanics of the junction, i.e., relationship between indentation force, tip bias, penetration depth, indentor geometry, and materials properties, ultimately determines the information obtained in the PFM experiment. This problem can be mapped on the electromechanics of the indentation of piezoelectric material, as discussed below.

**2. Dynamic behavior of the cantilever.**

Detected in most commercial SPM systems are amplitude and phase flexural and torsional oscillations of the cantilever used as a force or displacement sensor. Cantilevers have a complex frequency dependent dynamics, which can be both exploited to achieve higher signal/noise ratios (imaging at resonances),[47,48] or can significantly complicate quantitative data acquisition. A significant factor in PFM imaging is electrostatic tip-surface forces and buckling oscillations of the cantilever that can provide significant and in some cases even dominating contributions to the PFM signal.[49,50,51] Finally, electromechanical response is in general a vector having three independent components. While normal and lateral components can be determined from deflection and torsion of the cantilever, the use of the cantilever coupled with a beam-deflection detection system does not allow longitudinal force component along the cantilever axis to be unambiguously distinguished.[52] Frequency-dependent measurements of vector electromechanical response necessitate the relative magnitudes of vertical, lateral, and longitudinal responses to be calibrated.

**3. Electroelastic field structure inside the material.**

Phenomena such as resolution, tip-induced polarization switching, and PFM spectroscopy require the knowledge of electroelastic fields inside the material to analyze the thermodynamics and ultimately kinetics of domain generation, etc.

Below, we briefly discuss the principles of PFM on piezoelectric materials, as well as recent results in contact mechanics of tip-surface junction, vector PFM imaging in the low frequency regime, frequency dependence of PFM contrast, and approaches for PFM calibration. Image formation mechanism in PFM is analyzed in detail for piezoelectric materials in Section III. Brief analysis for ferroelectric materials is presented in Section IV. A novel approach for local electromechanical characterization, Piezoelectric Nanoindentation (PNI), is summarized in Section V. Finally, some future opportunities and challenges for PFM are summarized in Section VI.

## II.1. Principles of PFM

PFM is based on the detection of the bias-induced piezoelectric surface deformation. The tip is brought into contact with the surface, and the piezoelectric response of the surface is detected as the first harmonic component, $A_{1\omega}$, of the tip deflection, $A = A_0 + A_{1\omega}\cos(\omega t + \varphi)$, during application of the periodic bias $V_{tip} = V_{dc} + V_{ac}\cos(\omega t)$ to the tip. The phase of the electromechanical response of the surface, $\varphi$, yields information on the polarization direction below the tip. For $c^-$ domains (polarization vector oriented normal to the surface and pointing downward) the application of a positive tip bias results in the expansion of the sample and surface oscillations are in phase with the tip voltage, $\varphi = 0$. For $c^+$ domains, $\varphi = 180°$. The piezoresponse amplitude, $A = A_{1\omega}/V_{ac}$, given in the units of nm/V, defines the local electromechanical activity of the surface. The difficulty in the acquisition of PFM data stems from non-negligible electrostatic interactions between the tip and the surface, as well as between the cantilever and the surface. In the general case, the measured piezoresponse amplitude can be written as $A = A_{el} + A_{piezo} + A_{nl}$, where $A_{el}$ is the electrostatic contribution, $A_{piezo}$ is the electromechanical contribution and $A_{nl}$ is the non-local contribution due to capacitive cantilever-surface interactions.[49,50,53] Quantitative PFM imaging requires $A_{piezo}$ to be maximized to achieve predominantly electromechanical contrast. Provided that the phase signal varies by 180° between domains of opposite polarities, PFM images can be conveniently represented as $A_{1\omega}\cos(\varphi)/V_{ac}$, where $A_{1\omega}$ is the amplitude of first harmonic of measured response [nm]. Experimentally, collected signal is the output of the lock-in amplifier, and we refer the experimental signal as $PR = aA_{1\omega}\cos(\varphi)/V_{ac}$, given in the units of [V], where $a$ is a calibration constant determined by the lock-in settings and sensitivity of the photodiode. In addition to the vertical displacement of the cantilever, torsion of the cantilever can be measured as well, thus allowing measurement of both vertical and lateral PFM signals. Note however, that magnitude of these signals cannot be compared directly and thus do not form components of vector; some approaches to calibrations are considered below.

## III. PFM on Piezoelectric Materials

Here, we discuss in detail the imaging mechanism of PFM on linear piezoelectric materials, for which switching or non-linear coupling phenomena are absent.

### III.1. Contact Mechanics of Piezoelectric Indentation

Traditionally, principles and physical underpinnings of SPM techniques can be conveniently understood using force-distance curves (Fig. 6 a). Depending on the tip-surface separation and dominant interactions, contact mode atomic force microscopy (AFM) imaging in the repulsive region of Van-der-Waals forces, Atomic Force Acoustic Microscopy (AFAM) and Ultrasonic Force Microscopy (UFM) in the elastic indentation regime, non-contact AFM imaging in the attractive region of VdW forces, and magnetic and electrostatic imaging, can be distinguished. In all cases, the dominant force contribution controls the SPM mechanism and the information acquired from the experiment.

Similar approach can be used for voltage modulation techniques such as PFM. However, here the system is described by two independent variables – tip-surface separation and tip bias, giving rise to force-distance-bias surface as depicted in Fig. 6 b. In the non-contact regime, the tip-surface forces are purely capacitive and shape of the surface is described by $F_{nc} = C_z^{'}(z)(V_{tip} - V_{surf})^2$, where $C_z^{'}(z)$ is tip surface capacitance gradient. Non-contact voltage modulation techniques such as Kelvin Probe Force Microscopy (KPFM) are sensitive to voltage derivative of the force, $\partial F_{nc}/\partial V_{tip}$. Given the known parabolic form of this dependence and use of nulling approach, this renders KPFM readily interpretable and relatively insensitive to topographic artifacts. As opposed to it, techniques such as Electrostatic Force Microscopy (EFM) are sensitive to distance derivative of capacitance, and presence of (unknown) $C_z'(z)$ term renders quantitative interpretation of EFM more challenging.

In the contact regime, the imaging mechanism of SPM is ultimately controlled by the shape of the force-distance-bias surface, i.e. $F_c = F_c(h, V_{tip})$, where $h$ is indentation depth. Image formation mechanism in various SPM can be related to the derivatives of this surface, e.g. in the small signal approximation PFM signal is given by $(\partial h/\partial V_{tip})_{F=const}$, AFAM signal is related to $(\partial h/\partial F)_{V=const}$, UFM signal is determined by $(\partial^2 h/\partial F^2)_{V=const}$, as illustrated in Fig. 6 b. Image formation mechanism in recently developed frequency mixing techniques (Heterodyne Electrostatic-Ultrasonic Force Microscopy)[54] is controlled by mixed derivatives of the force-distance-bias surface, $\partial^2 h/\partial F \, \partial V$. Therefore, the knowledge of functional dependence of $F_c(h, V_{tip})$ is the key element for the quantitative interpretation of SPM on piezo- and ferroelectric material.

The contact electromechanics of piezoelectric materials is, however, extremely complex problem. A number of simplified approaches have been suggested based on the Green's function method. In these models, electric field in the material is calculated using rigid dielectric model ignoring piezoelectric coupling, the strain distribution is then calculated using the constitutive equations $x_i = d_{ik} E_k$, and displacement field is calculated from strain/stress field using numerical methods[55] or appropriate Green's function for semi-infinite elastic solid.[56] These solutions are ideally suited for modeling PFM signal in spatially inhomogeneous systems (domain walls, etc); however, the validity of the approximations made has not yet been established.

The rigorous solution of piezoelectric indentation is available only for the case of transversally isotropic material.[57,58,59] Karapetian[60,61] has derived rigorous description of contact mechanics in terms of stiffness relations between applied force, $P$, and concentrated charge, $Q$, with indenter displacement, $w_0$, indenter potential, $\psi_0$, indenter geometry and materials properties. The solutions were obtained for flat, spherical, and conical indenter geometries, and have the following phenomenological structure:

$$P = \frac{2}{\pi}\theta\left(h^{n+1}C_1^* + (n+1)h^n\psi_0 C_3^*\right) \tag{9}$$

$$Q = \frac{2}{\pi}\theta\left(-h^{n+1}C_3^* + (n+1)h^n\psi_0 C_4^*\right) \tag{10}$$

where $h$ is total indenter displacement, $\theta$ is geometric factor [ $\theta = a$ for flat indenter, $\theta = (2/3)R^{1/2}$ for spherical indenters and $\theta = (1/\pi)\tan\alpha$ for conical indenter] and $n = 0$ for flat, $n = 1/2$ for the spherical and $n = 1$ for the conical indenters, respectively.

These stiffness relations provide an extension of the corresponding results of Hertzian mechanics and continuum electrostatics to the transversely isotropic piezoelectric medium. By comparing equation (9) with stiffness relations for isotropic elastic solid for three indenter geometries studied, the *indentation elastic stiffness* $C_1^*$ for the piezoelectric indentation problem is analogous to the effective Young's modulus for isotropic material, $E^* = E/\left(1 - \nu^2\right)$, where $E$ is Young's modulus of the material below the indenter and $\nu$ is Poisson's ratio. In the isotropic limit, $C_1^* = \pi E^*$. Similarly, by comparing the indenter charge (10) with the capacitance of the conductive disc on dielectric half-plane, *indentation dielectric constant* $C_4^*$ is found to be analogous to the dielectric constant for the uniform material, and in the isotropic limit $C_4^* = 2\pi\kappa$. Electromechanical coupling is determined by *indentation piezocoefficient* $C_3^*$. In contact problem, ratio $C_3^*/C_1^*$ describes coupling between the force and the charge and the potential and displacement, similarly to the $d_{33}$ in the uniform field case. All indentation stiffnesses are complex functions of electroelastic constants of material, $C_i^* = C_i^*(c_{ij}, e_{ij}, \varepsilon_{ij})$, where $c_{ij}$ are elastic stiffnesses, $e_{ij}$ are piezoelectric constants, and $\varepsilon_{ij}$ are dielectric constants. Detailed analysis of stiffness relations for the spherical indentation and effect of materials constants on values of coupling coefficients is given elsewhere.[60] It has been shown that for most materials $C_3^*/C_1^* \approx d_{33}$ and $C_4^* \approx \sqrt{\varepsilon_{11}\varepsilon_{33}}$, validating earlier approximations in the interpretation of PFM.

This analysis yields a number of important conclusions on the information that can be obtained from SPM or nanoindentation experiment on the transversally isotropic piezoelectric material (e.g., $c^+$, $c^-$ domains in tetragonal perovskites) characterized by ten independent electroelastic constants. For all simple tip geometries, materials properties are described by three parameters, indentation elastic stiffness, $C_1^*$; indentation piezocoefficient, $C_3^*$; and indentation dielectric constant, $C_4^*$. Thus, the maximum information on electroelastic properties for a transversally isotropic material that can be obtained from an SPM experiment is given by these three quantities and mapping of $C_i^*$ distributions provides a comprehensive image of surface electroelastic properties. Experimentally, AFAM and UFM response is determined by $C_1^*$, while PFM is sensitive to $C_3^*/C_1^*$. Due to the smallness of corresponding capacitance, indentation dielectric constant, $C_4^*$, cannot be directly determined in the SPM experiment; however, it might be accessible on the larger length scales e.g. using nanoindentation approach.

While the rigorous (and even approximate) description of piezoelectric indentation is not yet available for materials with lower symmetry, it can be conjectured that in analogy with indentation anisotropic elastic materials, Eqs. (9,10) will be valid for arbitrary materials. In addition, the in-plane component of surface displacement in this case will be non-zero. While exact or even approximate description of in-plane electromechanics is not available, the zero order approximation will be that components of surface displacement are given by the normal and shear elements of piezoelectric constant tensor, $(w_1, w_2, w_3) = (d_{34}, d_{35}, d_{33})$, as discussed in detail elsewhere.[62]

### III.2. Image Formation Mechanism at Low Frequencies

### III.2.1. Vertical PFM signal.

The contrast formation mechanism in PFM is determined by the interplay of contact mechanics of the tip-surface junction and cantilever dynamics. In the low frequency regime, the mechanical equivalent circuit can be represented by two springs, connected in series, having spring constants $k_1$ and $k_2$, as shown in Fig. 7. Local electromechanical contributions to the vertical PFM signal arise due to the bias induced surface displacement, represented as $d_1$ (vertical) and $d_2$ (longitudinal). Note that for the cantilever based force sensor, vertical and lateral contact mechanics are coupled, and even for a purely vertical PFM (VPFM) signal, the motion of the tip along the surface will result in change of deflection angle (Fig. 7 d), as will be discussed in Section III.2.2.

In the Hertzian approximation[63] for a spherical tip, the vertical tip-surface junction spring constant is $k_1 = 1.82 (E^*)^{2/3} R^{1/3} P^{1/3}$, where $P$ is indentation force, $R$ is tip radius of curvature and $E^*$ is the indentation modulus. The indentation force is $P = kA_0$, where $k$ is the spring constant of the cantilever and $A_0$ is the static set-point cantilever deflection. The indentation mechanics for piezoelectric materials is more complex, and an exact solution is available only for transversally isotropic piezoelectric materials.[60] For spherical tip, the spring constant is $k_1 = R^{1/2} \left( 2 w_0^{1/2} C_1^* - w_0^{-1/2} V_{tip} C_3^* \right) \Big/ \pi$, where $w_0$ is an indentation depth determined by the stiffness relation Eq. (9). For a typical ferroelectric, such as BaTiO$_3$, in the $c^+$ domain state, with an indentation elastic stiffness $C_1^* = 403 \, \text{GPa}$, an indentation piezoelectric stiffness $C_3^* = 15.4 \, \text{N/Vm}$, a tip radius of $R = 50$ nm, an applied force of $P = 100$ nN, the indentation depth is $w_0 = 3.01 \text{A}$, and the effective tip-surface spring constant is $k_1 = (993 - 63.3 V_{\text{tip}}) \text{N/m}$. The bias dependence of the tip-surface spring constant is relatively weak and becomes even smaller for a flattened tip. The contact spring constant, $k_1 \sim 1000 \, \text{N/m}$, is significantly higher than the typical cantilever spring constant $k \sim 1 - 50 \, \text{N/m}$. Thus, in the low frequency regime, the vertical tip displacement can be found as $\delta A = k_1 \delta w / (k + k_1)$ (Fig. 7b), where $\delta w = d_1 V_{ac}$ is the bias induced surface displacement. Hence, the tip deflection is almost equal to the surface displacement, $\delta A \approx \delta w$, which is the usual assumption in PFM.

### III.2.2. Lateral and longitudinal PFM mechanics.

In the general case of a piezoelectric sample with arbitrary crystallographic orientation, application of the bias to the tip results in the surface displacement, $\mathbf{w}$, with both normal and in-plane components, $\mathbf{w} = (w_1, w_2, w_3)$. It is generally agreed that the use of a conventional four-quadrant photodetector allows the lateral piezoresponse component in the direction normal to the cantilever axis (lateral transversal displacement) to be determined as torque of the cantilever. Thus, if the cantilever orientation is given by the vector $n = (\cos\theta_c, \sin\theta_c, 0)$, where $\theta_c$ is the angle between the long axis of the cantilever and $x$-axis of the laboratory coordinate system, the lateral PFM (LPFM) signal is proportional to the projection of the surface displacement on the vector perpendicular to the cantilever axis, $PR_p = b(-w_1 \sin\theta_c + w_2 \cos\theta_c)$. The fundamental difference between VPFM and LPFM is that in the latter case the displacement of the tip apex can be significantly smaller than that of the surface, e.g. because of the onset of sliding friction.[64] Another issue in the LPFM imaging is the presence of the piezoresponse component along the cantilever axis (longitudinal displacement), $PR_l = c(w_1 \cos\theta_c + w_2 \sin\theta_c)$. This longitudinal displacement couples to the vertical signal and can be determined from comparison of VPFM images obtained for different cantilever orientations in the X-Y plane as discussed below.

It is important to emphasize that the simple combination of VPFM and LPFM measurements is insufficient to unambiguously determine the 3D piezoresponse vector for an arbitrarily oriented sample. To overcome this limitation, sequential acquisition of two LPFM images at two orthogonal orientations of the sample with respect to the cantilever, further referred to as $x$-LPFM and $y$-LPFM images, has been accomplished by using the etched top electrodes as topographic markers.[65,66] During these measurements, the laboratory coordinate system is selected such as $\theta_c = 0$ for $x$-PFM and $\theta_c = \pi/2$ for $y$-PFM. Thus, the relationship between measured piezoresponse signals and the surface displacement vector is:

$$
\begin{pmatrix} xPR_v \\ xPR_l \\ yPR_v \\ yPR_l \end{pmatrix} = \begin{pmatrix} c & 0 & a \\ 0 & b & 0 \\ 0 & c & a \\ b & 0 & 0 \end{pmatrix} \begin{pmatrix} w_1 \\ w_2 \\ w_3 \end{pmatrix},
\tag{11}
$$

This analysis can be extended for non-orthogonal scan directions in a straightforward manner. Eq. (11) allows the contribution of the longitudinal displacement to the VPFM signal to be determined from the ratio $\beta = (xPR_v - yPR_v)/(xPR_v + yPR_v)$, spatial map of which allows the contribution of longitudinal surface displacement to VPFM signal to be determined. If $\beta \ll 1$ within the image, VPFM signal is artifact free.

In the case, when vertical PFM does not contain a significant contribution from the longitudinal surface displacement $x$-VPFM and $y$-VPFM images are identical, $xPR_v = yPR_v = vPR$ and Eq. (11) becomes

$$
\begin{pmatrix} xPR_l \\ yPR_l \\ vPR \end{pmatrix} = \begin{pmatrix} 0 & b & 0 \\ b & 0 & 0 \\ 0 & 0 & a \end{pmatrix} \begin{pmatrix} w_1 \\ w_2 \\ w_3 \end{pmatrix},
\tag{12}
$$

Eq. (12) contains two independent calibration constants, $a$ and $b$. The calibration constant for the VPFM signal, $a$, can be determined in a straightforward way by using an external reference, e.g., piezoelectric sample with well known piezoelectric constants, such as quartz, in the integral excitation (metal-coated top surface) configuration.[67,68,69] Alternatively, the sample can be mounted on a calibrated piezoelectric transducer and surface vibration at low frequencies below the cantilever and transducer resonances can be used to calibrate the tip oscillation amplitude. A similar approach can be used for lateral[69] and longitudinal calibration, as analyzed in section III.4.

### III.2.3. Vector PFM.

An example of 2D PFM image is illustrated in Fig. 8, which shows vertical and lateral PFM images of PMN-PT single crystal. Vertical PFM image illustrates the presence of antiparallel $c$ domains, while lateral image shows the response at the domain walls because of the tilting of the surface. To represent vector PFM data, the VPFM and LPFM images are normalized so that the intensity changes between -1 and 1, i.e. $vpr, lpr \in (-1,1)$. Using commercial software,[70] 2D vector data $(vpr, lpr)$ is converted to the amplitude/angle pair, $A_{2D} = \text{Abs}(vpr + I \, lpr)$, $\theta_{2D} = \text{Arg}(vpr + I \, lpr)$. This information can be represented using vector image, where the color corresponds to the orientation, while intensity corresponds to the magnitude. Alternatively, this data can be represented in the scalar form by plotting separately phase $\theta_{2D}$, and magnitude, $A_{2D}$, as illustrated in Figs. 8d and e, respectively. Note that two types of domain walls can be observed on the amplitude image – "bright" walls parallel to the cantilever axis at which the electromechanical activity of the surface is enhanced, and "dark" walls perpendicular to the cantilever axis at which electromechanical activity is decreased. This asymmetry is due to the difference in signal transduction between longitudinal and lateral response components, as described above.

### III.2.4. Resolution in PFM.

One of the crucial parameters in any microscopy is the spatial resolution. In PFM, numerous reports of imaging with sub-10 nanometer resolution are available; however, in most cases resolution is not well defined and is identified with either characteristic domain wall width or minimal feature size that can still be observed in the image. An approach to unambiguous definition of minimal feature size is shown in Fig. 9 illustrating the checkerboard domain pattern written on the lead zirconate titanate (PZT) thin-film surface and corresponding 2D Fourier Transform (FT). The distance corresponding to the still visible reflex with largest wavevector defines the minimum observable feature size. This behavior is illustrated in Fig. 9c, plotting the wavevector dependence of the Fourier peak intensity. The minimum feature size is limited by the noise level of the system. As opposed to it, resolution can be formally determined from the ratio of the Fourier intensities for the ideal and theoretical image and depends only on the probe characteristics (Fig. 9d), as reported elsewhere. An alternative approach to the Fourier method is the use of the meshes with variable grid size, as illustrated in Fig. 9e,f. The minimum visible feature is ~30 nm, i.e., comparable to Fig. 9c. However, in this case the minimum feature size is not determined unambiguously.

### III.3. Frequency Dependence of PFM Contrast

### III.3.1. Cantilever dynamics.

A cantilever in combination with an optical beam deflection detector is the key part of the SPM force detection mechanism. The motion of the cantilever induced by surface oscillations has been studied extensively in the context of AFAM[71, 72, 73, 74] and UFM.[75] However, electrostatic modulation in PFM gives rise to additional local and non-local force contributions that can couple to the displacement induced oscillations. The analysis of the dynamic image formation mechanisms in vector PFM should necessarily take into account the following contributions as illustrated in Fig. 7 a,c,d:

1. The local vertical surface displacement translated to the tip.

2. The longitudinal, in-plane surface displacement along the cantilever axis.

3. The lateral surface displacement, in-plane and perpendicular to the cantilever axis.

4. The local electrostatic force acting on the tip.

5. The distributed electrostatic force acting on the cantilever.

The basic features of the dynamic behavior of the cantilever can be described by the beam equation

$$\frac{d^4u}{dx^4} + \frac{\rho S_c}{EI}\frac{d^2u}{dt^2} = \frac{q(x,t)}{EI}, \tag{13}$$

where $E$ is the Young's modulus of cantilever material, $I$ is the moment of inertia of the cross-section, $\rho$ is density, $S_c$ is cross-section area, and $q(x,t)$ is the distributed force acting on the cantilever. For a rectangular cantilever $S_c = wh$ and $I = wh^3/12$, where $w$ is the cantilever width and $h$ is thickness. The cantilever spring constant, $k$, is related to the geometric parameters of the cantilever by $k = 3EI/L^3 = Ewh^3/4L^3$. In beam-deflection SPM, the deflection angle of the cantilever, $\theta$, is measured by the deflection of the laser beam at $x = L$ and is related to the local slope as $\theta = \arctan\left[u_0^{'}(L)\right] \approx u_0^{'}(L)$. For a purely vertical displacement, the relationship between cantilever deflection angle and measured height is $A = 2\theta L/3$.[76] Thus, in cases when the deflection angle is determined by either longitudinal or electrostatic contributions, the effective vertical displacement measured by AFM electronics will also be related to the deflection angle as $A = 2\theta L/3$.

Eq. (13) is solved in the frequency domain by introducing $u(x,t) = u_0(x)e^{i\omega t}$, $q(x,t) = q_0 e^{i\omega t}$, where $u_0$ is the displacement amplitude, $q_0$ is a uniform load per unit length, $t$ is time, and $\omega$ is modulation frequency. After substitution, Eq. (13) is:

$$\frac{d^4 u_0}{dx^4} = \kappa^4 u_0 + \frac{q_0}{EI},$$ (14)

where $\kappa^4 = \omega^2 \rho S_c / EI$. On the clamped end of the cantilever, the displacement and deflection angle are zero, yielding the boundary conditions

$$u_0(0) = 0 \text{ and } u_0^{'}(0) = 0,$$ (15a,b)

On the supported end, in the limit of linear elastic contact the boundary conditions for moment and shear force are

$$EIu_0^{''}(L) = k_2 H\left(\tilde{d}_2 - u_0^{'}(L)H\right) \text{ and } EIu_0^{'''}(L) = -f_0 + k_1\left(u_0(L) - d_1\right)$$ (16a,b)

where is $d_1 = d_{vert}V_{ac}$ is the first harmonic component of bias-induced vertical surface displacement due to the piezoelectric effect, $d_2 = d_{lat}V_{ac}$ is the first harmonic component of the longitudinal surface displacement, $f_0$ is the first harmonic of the local force, $f(x,t) = f_0 e^{i\omega t}$, acting on the tip, and $k_1$ and $k_2$ are the vertical and longitudinal spring constants of the tip-surface junction (Fig. 7a). For non-piezoelectric materials, $d_1 = d_2 = 0$, while for zero electrostatic force, $f_0 = 0$, providing purely electromechanical and purely electrostatic limiting cases for Eq. (14).

Because Eq. (14) is linear, it can be solved in the usual fashion. Using $EI = kL^3/3$, the dynamic behavior of the cantilever is given by

$$\theta_{tot} = \frac{A_v(\beta)d_1 + A_l(\beta)d_2 + A_e(\beta)f_0 + A_q(\beta)q_0}{N(\beta)}$$ (17)

where

$$A_v(\beta) = 3\beta^4 k_1 kL \sin\beta \sinh\beta \qquad (18)$$

$$A_l(\beta) = 3\beta^2 H k_2 \left[3k_1 + \cosh\beta\left(-3k_1\cos\beta + \beta^3 k \sin\beta\right) + \beta^3 k \cos\beta \sinh\beta\right] \qquad (19)$$

$$A_e(\beta) = 3\beta^4 kL \sin\beta \sinh\beta \qquad (20)$$

$$A_q(\beta) = 3L^2\left[3k_1(\cos\beta - \cosh\beta) - k\beta^3 \sin\beta + \left(k\beta^3 + 3k_1\sin\beta\right)\sinh\beta\right] \qquad (21)$$

$$N(\beta) = \beta^2[9H^2 k_1 k_2 + \beta^4 k^2 L^2 + \cosh\beta((-9H^2 k_1 k_2 + \beta^4 k^2 L^2)\cos\beta + \\ + 3\beta k(k_1 L^2 + H^2 k_2 \beta^2)\sin\beta) + 3\beta k(-k_1 L^2 + H^2 k_2 \beta^2)\cos\beta\sinh\beta] \qquad (22)$$

and the dimensionless frequency is $\beta = \kappa L$.

The ratios $A_v(\beta)/N(\beta)$, $A_l(\beta)/N(\beta)$, $A_e(\beta)/N(\beta)$, and $A_q(\beta)/N(\beta)$ describe the frequency dependence of the PFM signal due to vertical and longitudinal components of surface displacement, the local electrostatic force acting on the tip, and the distributed electrostatic force acting along the cantilever, respectively. Note that the vertical electromechanical contribution and local force contribution have similar frequency dependences [compare Eqs. (18) and (20)].

The resonance structure in Eq. (17) is determined only by the properties of the cantilever and the spring constant of the tip-surface junction and is independent of the relative contributions of electrostatic and electromechanical interactions. Therefore, tracking the resonant frequency as a function of tip position provides information on local elastic properties, which is similar to frequency detection in AFAM.[77] Since the denominator of Eq. (17) does not depend on the relative magnitudes of vertical, longitudinal, and electrostatic responses, these contributions cannot be separated by a proper choice of driving frequency. Therefore, unambiguous measurement of all three components of the electromechanical response vector requires alternative solutions, e.g., based on either 3D SPM[78] or sample rotation.[65,66] At the same time, electromechanical, local, and non-local contributions to the PFM signal are additive, making it possible to distinguish the relative contributions of these signals to the observed contrast.

From Eqs. (18-22), the frequency dependence of the non-local electrostatic, local electrostatic, and piezoelectric contributions to be estimated as $A_v(\beta)/N(\beta) \sim k_1/\omega$, $A_l(\beta)/N(\beta) \sim k_1/\omega^{3/2}$, $A_e(\beta)/N(\beta) \sim 1/\omega$, and $A_q(\beta)/N(\beta) \sim 1/\omega^{3/2}$. All four contributions decrease with frequency because of the dynamic stiffening effects. Even in the absence of damping, the non-local contribution scales as a higher power of frequency, suggesting that non-local cantilever effects will be minimized at high frequencies. At the same time, the local electrostatic and electromechanical contributions scale in a similar manner as the ratio, $A_{piezo}/A_{el} = d_1 k_1/f_0$, which depends only on the spring constant of the tip-surface junction. This suggests that these contributions cannot be distinguished by a choice of the operating frequency. Instead, either the use of a cantilever with a high spring constant ($k_1 \to \infty$) or imaging at the nulling bias or using shielded probes[79] (where $f_0 \to 0$) is required. The detailed analysis of frequency dynamics in PFM is given elsewhere.[80]

### III.3.2. Response maps.

Eq. (17) predicts complex frequency dynamics of the PFM, in which relative contribution of electrostatic and electromechanical contrasts strongly depend on frequency. This behavior can be represented in the form of response maps in Fig. 10 as a function of frequency and tip-surface potential difference, $\Delta V$, and calculated according to Eq. (28) for zero local electrostatic force, $f = 0$. A number of resonances (bright lines) and antiresonances (black lines) can be clearly seen. The phase changes by 180° across resonance and antiresonance lines. For low tip biases, the response is purely electromechanical and is independent of ΔV. For higher dc biases, the response is dominated by non-local contributions and is linear in ΔV. Note that the position of the resonances is determined solely by the cantilever properties and spring constant of the tip-surface junction and is independent of tip bias. Thus, the resonance frequency of the electrically driven cantilever in contact with the surface provides information only on the elastic properties of material, but not piezoelectric properties. At the same time, the zeroes on the response diagram are strongly bias dependent and therefore, the magnitude and frequency dependence of the nulling bias is related to the magnitude and sign of the electromechanical response. The relative magnitudes of non-local and electromechanical contributions are illustrated in Fig. 10 b,d,f, illustrating the response map for $A_{piezo}/\left(A_{piezo} + A_{nl}\right)$. The white region corresponds to dominant electromechanical contrast, while black regions correspond to dominant non-local electrostatic contributions. Note that in the low frequency limit, the crossover between the two (indicated by an arrow) scales proportionally to the cantilever spring constant. At high frequencies, the relative contribution of the electromechanical contrast increases, indicative of cantilever stiffening. Also note that in the vicinity of the antiresonances the non-local contribution is enhanced, while the resonances do not affect the relative contributions of these signals. Therefore, imaging at cantilever resonances will increase the signal to noise ratio, but will not affect the relative contributions of electrostatic and electromechanical responses, thus justifying the applicability of contact resonance-enhanced PFM imaging for low coercive bias materials.

### III.3.3. 2D Force-bias spectroscopy.

Experimentally, frequency, and bias-dependent dynamics of the cantilever can be accessed using 2D spectroscopy, in which local electromechanical response is measured as a function of frequency and dc bias offset on the tip. Shown in Fig. 11 a,b is the vertical response amplitude for non-piezoelectric $SiO_2$ in non-contact and contact modes. The three major non-contact resonances at $\omega_1 = 54.5$ kHz, $\omega_2 = 346.7$ kHz, and $\omega_3 = 980.5$ kHz are clearly seen. The ratio of the frequencies is $\omega_3 : \omega_2 : \omega_1 = 17.97 : 6.35 : 1$, very close to the theoretical ratio $17.55 : 6.23 : 1$. The frequency-independent nulling bias is $V_{tip} = 1.0$ V. Response diagrams for the contact regime is shown in Fig. 11 b. The resonances in the contact regime are $\omega_1 = 407.9$ kHz and $\omega_2 = 1075.0$ kHz. The ratio of the resonant frequencies is $\omega_2 : \omega_1 = 2.63 : 1$, as compared with the theoretical ratio of 3.24:1. An additional resonance at $\omega = 634.1$ kHz emerges that can be attributed to rotation of the cantilever with respect to tip-surface junction.

For $SiO_2$, the nulling bias is frequency independent in the contact regime. A similar response diagram measured for PZT in Fig. 11 c,d shows a completely different behavior. The nulling bias is now strongly frequency dependent, as expected for the case when the relative contributions of electrostatic and electromechanical signals vary due to different frequency dependence (comp. Fig. 10 a,c,e). The frequency dependence of nulling bias depends on grain orientation, as shown in Fig. 11 c,d. Note that the orientation of the line corresponding to the frequency dependence of the nulling bias (vertical dark line) is opposite for these two grains, indicative of the opposite signs of the electromechanical contribution to the PFM signal. This behavior is further illustrated in Fig. 12 a for the frequency dependence of vertical and lateral PFM signals. Note that resonance frequencies in vertical PFM are material dependent, as determined by the difference in elastic properties. At the same time, lateral response decreases rapidly above 10 kHz, indicative of the onset of tip sliding along the surface. The subsequent increase of the lateral signal above 100 kHz is due to capacitive cross-talk in AFM electronics.

This analysis allows the frequency range for optimal PFM imaging to be established using the deviation of nulling bias from surface potential value as a measure of electromechanical contribution to the signal (Fig. 12 b). Indeed, for purely electrostatic imaging, the nulling bias is equal to the surface potential. In the electromechanical limit, the response is bias independent, and there is no nulling bias. In the intermediate case, the nulling bias is $V_{null} = V_{surf} \pm d_{eff} / G_{elec}$ (the sign corresponds to the domain orientation) and depends on the relative magnitudes of electrostatic and electromechnical contributions. Therefore, the frequencies for which $|V_{null}|$ is maximal correspond to the frequencies at which the electromechanical contribution is dominant, and the resulting PFM image has optimal contrast.

### III.3.4. Deconvolution of electrostatic and electromechanical contributions.

This analysis can be extended to deconvolute electromechanical and electrostatic contributions to the PFM signal to be distinguished. As discussed above, the PFM $x$-signal, defined as $PR = A\cos\varphi$, can be represented as

$$PR_+ = \tilde{d}_{eff} + \tilde{G}_{elec}\left(V_{dc} - V_{surf}\right), \tag{23}$$

where $\tilde{d}_{eff}$ and $\tilde{G}_{elec}$ are the electromechanical and electrostatic contributions now including a frequency-dependent phase multiplier. From Eq. (23), the electrostatic contribution to the PFM signal can be determined from the slope, $\tilde{G}_{elec} = c$, of the response vs. bias curve at each frequency. The electromechanical contribution is related to the intercept, $b$, as $\tilde{d}_{eff} = b + cV_{surf}$. Note that while the electrostatic contribution can be determined unambiguously, the electromechanical contribution depends on a known surface potential, $V_{surf}$, which can be determined e.g., from non-contact measurements.

To illustrate applicability of this approach, we extend this analysis to the data in Fig. 11 e. The function $y = b + cV_{tip}$ was fit to the signal for each frequency. Shown in Fig. 11 f is error map, $PR - \left(b + cV_{tip}\right)$, representing the deviation of the actual response from a purely linear response. The scale for Fig. 11 f is 1% of full scale for Fig. 11 e. Note that the deviation from linearity is extremely small, suggesting the validity of Eq. (23).

The frequency dependences of the electromechanical and electrostatic responses for these materials are shown in Fig. 12 c,d. Note that the electromechanical response is greatest for PZT grain 1 with a good tip, slightly smaller for grain 2, and is negligibly small for $SiO_2$ in the non-contact and contact modes (Fig. 12 a), as expected. In comparison, the electrostatic response is comparable for all materials (Fig. 12 b). The resonant frequencies for the electromechanical and electrostatic signals coincide for a given sample, as predicted by Eq. (17). An alternative approach for distinguishing electrostatic and electromechanical contributions has been suggested by Harnagea[48] based on measurements of an amplitude-frequency curve of piezoelectric and non-piezoelectric materials. However, this approach is applicable only if the resonant frequencies of a cantilever in contact with the surface are identical, which is not the case for dissimilar materials. The differentiation of these contributions based on the bias dependence of the response provides a more rigorous approach provided that the surface potential is known.

**III.4. Calibration and Cantilever Dynamics in PFM**

Quantitative interpretation of PFM experiments necessitates the studies of the oscillation transfer mechanism between the piezoelectric surface and the SPM tip. A convenient approach for such calibration is based on macroscopic piezoelectric actuators producing vertical and in-plane oscillations of known amplitude and direction.[69] A series of measurements were performed in order to establish quantitative information relating motion induced at the probe tip to the measured response at the AFM photodetector. The experimental set-up is shown in Fig. 13. Either a normal or shear mode piezo actuator was placed between the probe tip and the sample stage and voltage signals could be applied to either the actuator or the tip independently. Within this set-up, ten different experimental scenarios were explored in which the direction of piezo oscillation, the mode of cantilever deflection measured (vertical or torsional), the proximity of the tip to the surface, and the driving signal connection scheme were varied. The 2D-spectroscopic results of these experiments are plotted in Fig. 14, for a cantilever with a 4.5 N/m spring constant, and Fig. 15 a (40 N/m) and Fig. 15 b (0.15 N/m) and labeled accordingly:

- *v* or *l* refers to whether the *v*ertical or *l*ateral photo-detector signal is plotted.

- *c* or *n* refers to whether the probe tip is in direct *c*ontact with the surface, or *n*ot-in-contact, just above the surface.

- *z*, *x*, and *y* refer to the oscillation direction of the piezo-actuator in accordance with Fig. 13. *z* induced vertical oscillation, *x* induced longitudinal shear, and *y* induces lateral shear.

- *1* and *2* refer to the signal connection scheme. *1* corresponds to Fig. 13b (ac bias applied to the tip, electrostatic excitation) and *2* to Fig. 13a (ac bias applied to the actuator, mechanical excitation).

The presence of a large number of resonances within the piezo-actuators themselves contributes to the large number of resonances apparent when the tip is in contact with the surface. However, several prominent and important features are visible.

Comparison of *vnz1* and *lnz1*, the measured vertical and lateral responses when the tip is not in contact with the surface and driven exclusively by long-range electrostatic interactions, shows relatively featureless spectra, with the primary free-resonance peaks being the largest (1). Cross-talk between vertical and lateral measurements accounts for the appearance of the vertical resonant peak in the lateral spectrum. The secondary resonant peak (2) is visible in the *vnz1* amplitude and phase plots, but is hardly visible in the *lnz1* amplitude plot and detectable only as a 180º change in phase in Fig. 14 b (3). We have consistently found vertical/lateral cross-talk to diminish at higher frequencies ( >1 MHz). The nulling potential, visible as a vertical line of smallest response (4), remains constant as a function of frequency indicating the absence of any piezoelectric effects.

A similar analysis of *vcz1* and *lcz1*, measured with the tip in contact with the surface, shows that the primary resonance has been shifted upwards as a result of the boundary constraints as compared with the non-contact case above and that the cross talk between vertical and lateral signals is small at this frequency. A number of unidentifiable resonances are visible but can be distinguished from the cantilever resonance by the existence of the nulling potential (5).

Measurements made when the piezo actuator was driven show very complex behavior. In this case, no nulling potential is observed, since the signal is dominated by mechanical forces. While detailed interpretation is complex, several prominent features can be delineated. In the case of vertical oscillator ($vcz2$, $lcz2$), both vertical and lateral signal are observed either because of the cross-talk in the photodiode or presence of local shear component of surface wave. However, the intensity of lateral signal decreases compared to vertical signal at high frequencies, indicative of the onset of sliding friction. Similar behavior is observed for the lateral oscillator ($vcy2$, $lcy2$), in which case lateral signal is expected to be dominant and vertical is a crosstalk. Most notably, for longitudinal surface oscillations ($vcx2$, $lcx2$), both vertical and lateral signals are non-zero (albeit weaker then primary responses) and demonstrate similar dynamics.

To study the cantilever effect on the measurements, shown in Fig. 15 are the results of measurements of cantilevers with much higher (40 N/m) and lower (0.15 N/m) spring constants. In the former case, only the first contact resonance falls into the frequency range of study. Thus, the data in Fig. 15 a is collected in the low-frequency regime and represents the intrinsic oscillator responses. On the contrary, shown in Fig. 15 b is the response diagram for soft cantilever with significant number of resonances in the frequency range of study. However, despite the fact that there is a correspondence between resonances under electrostatic and mechanical excitations, the number of the latter is significantly higher, indicative of the dominant role of oscillator dynamics on the system response.

To summarize, piezoelectric actuators provide an approach for calibration of the frequency-dependent response of the PFM and address the coupling between three components of surface response vector and measured vertical and lateral PFM signal.

However, this approach necessitates the oscillators with known (ideally frequency independent) response. These can be developed using microactuators, as reported elsewhere.

## IV. PFM on Ferroelectric Materials

In ferroelectric materials, local polarization can be switched by the application of electric or mechanical stimuli. In the context of SPM, this opens a pathway to modify local domain structures using electrostatic and mechanical action of the tip, for applications such as spectroscopy and domain patterning. Several recent and forthcoming papers and reviews summarize the progress in this field in great detail.[34,35,81,82,83] Here, we briefly summarize several prominent directions for PFM imaging and manipulations of ferroelectric materials.

### IV.1. Domain Patterning and Control

Application of a relatively small dc voltage between the tip and bottom electrode generates an electric field of several hundred kilovolts per centimeter, which is higher than the coercive voltage of most of ferroelectrics, thus inducing local polarization reversal. The driving force for the 180° polarization switching process in ferroelectrics is change in the bulk free energy density,[84, 85] $\Delta g_{bulk} = -\Delta P_i E_i - \Delta d_{i\mu} E_i X_\mu$, where $P_i$, $E_i$, $X_\mu$, and $d_{i\mu}$, are components of the polarization, electric field, stress and piezoelectric constants tensor, correspondingly, $i = 1,2,3$, and $\mu = 1,..,6$. The first and second terms describe ferroelectric and ferroelectroelastic switching, respectively.

The thermodynamics of the switching process can be understood from the analysis of the free energy of nucleating domain comprising the contributions form bulk free energy, domain wall energy, and depolarization field energy as $\Delta G = \Delta G_{bulk} + \Delta G_{wall} + \Delta G_{dep}$. The analysis of the thermodynamics of domain switching in the Landauer approximation[86] for domain shape and point charge approximation for the tip was given by Abplanalp[84] and independently by Molotskii et al.[87, 88] It was shown by Kalinin et al., that this model is limited only for large domain sizes, while the description of switching on the length scales comparable to the tip radius of curvature and higher order switching phenomena,[85] requires exact electroelastic field structure to be taken into account. For realistic tip geometries, domain nucleation requires certain threshold bias of order of 0.1-10 V,[81,84,,89] required to nucleate domain in the finite electric field of the tip.

In the last several years, it has been demonstrated that domain size in the PFM switching process is ultimately controlled by the kinetics of the switching process. A number of experimental and theoretical studies of domain growth kinetics have been reported.[39,90,91] It was shown that in general domain growth follows approximately logarithmic dependence on pulse length and linear dependence in magnitude. Several attempts to interpret this behavior in terms of the activation field for the domain wall motion and dynamics of the systems with frozen disorder have been made; however, the use of simplified point-charge like models that do not include the effects of the spherical and conical parts of the tip necessitates further theoretical and experimental studies of these phenomena.

## IV.2. PFM Spectroscopy

One of the most important applications of PFM includes local spectroscopy, i.e., measurements of local vertical and lateral hysteresis loops at the ~10 nm level.[36, 92] PFM hysteresis loops readily provide information on relative electromechanical activity and switching bias variation between dissimilar grains. Quantitative interpretation of PFM spectroscopy presents a complex problem due to the inhomogeneous distribution of the tip-generated field and random grain orientation.[93, 94,95] Based on the 1D model by Ganpule et al.,[96] a simplified description for the hysteresis loop shape was derived as $PR = \alpha d_{33}\left(1 - \eta/V_{dc}\right)$, where $PR$ is the piezoresponse amplitude, $PR = A_{piezo}/V_{ax}$ and $\eta$ is parameter determined by materials properties.[95] Recently, Kholkin et al.[97] has experimentally studied domain switching processes visualizing growing domain at each step of the process, and have demonstrated evolution of domain shape and the role of grain boundaries in switching process.

## IV.3. Non-Linear and Pressure Induced Phenomena in PFM

Application of relatively small biases or pressures to the SPM tip results in extremely electric fields and stresses due to the smallness of the contact area. This behavior often results in tip degradation during imaging, including peel-off of conductive coating and tip flattening. However, these extremely high fields in the vicinity of the tip potentially allow non-linear effects in ferroelectric and relaxor ferroelectric materials, such as pressure dependent piezoelectric properties and phase transitions, etc. to be studied. In these cases, changes in local electromechanical coupling due to the ferroelectric-antiferroelectric or ferroelectric-paraelectric phase transition or domain reorientation are detected as a function of indentation force. Load dependence of piezoresponse was originally reported by Zavala.[98] Abplanalp et al.[85] reported the high-order ferroelectroelastic switching in PFM. Recently, Kholkin et al.[99] has studied the pressure dependence of PFM signal for relaxor ferroelectrics using combination of PFM and interferometric techniques and reported observation of non-linear piezoelectric coupling phenomena.

However, despite this progress, applications of PFM to quantitative studies of ferroelectric materials beyond imaging are limited by the lack of quantitativeness and extreme sensitivity to the surface state of material, when the presence of minute contamination layer can result in the decay of the response.[100] Moreover, the presence of water layer inevitable in ambience strongly affects charge behavior on ferroelectric surfaces and potentially can affect PFM mechanism and domain dynamics,[101,102] especially at small loads required for high resolution imaging. These ambient effects can be minimized in the high- and ultrahigh vacuum environment; however, despite a number of attempts,[103,104] no systematic studies of the PFM in ultra high vacuum have been performed to date. An alternative approach for quantitative electromechanical measurements in the elastic and plastic regimes at the mesoscopic length scales, Piezoelectric Nanoindentation, is discussed below.

## V. Piezoelectric Nanoindentation

Nanoindentation is a popular quantitative tool for characterization of mechanical properties of materials on the nanoscale. The method itself was developed 25 years ago[105,106] and currently is well-established method of determination of mechanical properties of the materials at micro and nanoscale.[107] The method is based on accurate determination of applied load and resulting displacement of the indenter with known geometry into the sample. The data analysis is based on the Oliver-Pharr method.[108] In the most advanced form the method, uses continuous stiffness option when the indenter is additionally modulated with small periodical force, and the resulting periodical displacements of the indenter is measured with lock-in amplifier providing stiffness measurements during the experimental cycle.[107]

The multiple existing and emerging applications of electromechanically active materials have necessitated the development of quantitative tools for piezoelectric characterization on the nanoscale. A number of approaches based on the direct piezoelectric effect have been suggested, in which the current generated at the indenter-surface junction is collected during the experiment.[109,110,111,112,113] However, the total amount of generated charge is limited, resulting in low signal-noise ratios. Noteworthy, this approach cannot be extended to SPM because of the smallness of the tip-surface contact area.

Here, we discuss the electromechanical nanoindentation measurements based on the inverse piezoelectric effect, further referred to as Piezoelectric Nanoindentation (PNI). In PNI, periodic electric bias, $V = V_0 \cos(\omega t)$, is applied between indenter and bottom electrode, and the amplitude and phase of harmonic displacement are detected. Thus, PNI is an analog of the PFM implemented on the nanoindenter platform. The main benefit of this approach is independence of measurements from the load rate and using existing software and hardware without any external devises. The main advantage of PNI as compared to PFM is the capability to provide quantitative data, enabled by the calibrated displacement sensor, and capability to study electromechanical phenomena in a much broader range of loads. These techniques are compared in Table II.

PNI was implemented on a modified commercial nanoindenter (Nano XP MTS Corp.) (Fig. 16 a). A conducting nanoindentation tip was fabricated from electrochemically etched W wire and was connected to the ground during the experiment. The sample was mounted on an electrode that was subjected to an oscillating voltage. The experiment was controlled and data were collected and analyzed using TestWorks ™ software. The indenter displacement (penetration depth) and amplitude and phase of the piezoresponse were measured continuously during indenter approach to the surface, loading, holding at maximum load, and unloading.

Table 2. Relative characteristics of PFM and PNI

| Parameter | PNI | PFM |
|---|---|---|
| Quantitativeness | Yes | Semi; requires complex calibration |
| Lateral Resolution | ~100 nm | 5-10 nm |
| Vector response measurements | Yes; quantitative 3D nanoindentation has been reported | Only two components vertical and longitudinal coupled |
| Pressure-induced phenomena | Yes, including measurements with shear component of the load. | Difficult |
| Imaging mode | Slow | Yes |
| Working in liquid environment | Yes | Difficult |
| Variable temperature measurements | Yes | Yes |

Similarly to PFM, PNI data can be interpreted using the theory for elastic piezoelectric indentation developed by Karapetian et al.[61] Eq. (9). The PNI experiment is performed with load control of the indentation, and the changes in the load are slow in comparison with applied frequency. In the small signal limit, the measured piezoresponse amplitude is related to the amplitude of applied voltage $V_0$ as $V_0 \left| \left( \partial h / \partial E \right)_{P=const} \right|$ and from Eq. (9) piezoresponse amplitude is $PR = C_3^* / C_1^*$.

In the case of pure elastic contact and when the coupling constants are independent of the applied load, the measured piezoresponse is predicted to be independent of the indentation depth, a marked difference from dynamic contact stiffness which scales as the square root of the contact area. Here, we illustrate that this behavior holds for PZT piezoceramic, even when the contact is not purely elastic. The typical experimental results are shown in Fig. 16. The amplitude of sample voltage in this experiment was 1.5 V, and the applied frequency was 400 Hz. The load – unload curve is shown in Fig. 16 b, and the applied load as function of the time after initial contact is shown in Fig. 16 c. The fact that the loading and unloading curves are different indicates that some permanent deformation has occurred in the ceramics. The harmonic displacement rapidly increases after first contact but then remains essentially constant during all further loading, hold and unloading. The measured piezoresponse signal is 0.2 nm/V, as compared to an estimated average value of 0.3 nm/V for PZT.

Similar experiments performed at BaTiO$_3$ are shown in Fig. 17 a,b. The load – unload curves indicate again that some permanent deformation occur in this material. The piezoresponse rapidly increases at the moment of first indenter contact to about 1 mN to the value of 0.1 nm/V, but then decreases to value about 0.04 nm/V. During unloading there is a slight increase in harmonic displacement. The peak at load at about 0.5 mN may be attributed to resonance on the indenter / surface interaction, and the behavior at higher load is related with the onset of ferroelastic domain switching below the indenter tip, with associated decrease of electromechanical response, as reported elsewhere.[114]

Finally, PNI can be used for real space imaging of dynamics phenomena in ferroelectric materials. Shown in Fig. 17 c,d are the results of 20X20 mapping of $BaTiO_3$ sample with 2 micron step. The maps are acquired at the maximal and 10% loads, demonstrating the evolution of electromechanical response as a function of loading. Thus, PNI can be used both to visualize the domain patterns and follow the evolution the domain structures under mechanical stimuli.

## VI. Future Directions in PFM

In the decade since its invention, PFM has evolved from a relatively obscure and controversial technique for domain imaging to a broadly accepted tool for nanoscale characterization, domain patterning, and spectroscopy of ferroelectric material. PFM is applicable for imaging of other piezoelectric functional materials such as III-V nitrides, providing information on surface termination. Recently demonstrated potential of PFM for sub-10 nm imaging of electromechanically active proteins in calcified and connective tissues[44,45,46] suggests a much broader field of applications in the biological community, where the techniques for studies of ultrastructure of tissues on the sub-micron level are scarce. In this section, we summarize some of the unresolved challenges and opportunities in PFM.

### VI.1. Advanced PFM Probes

One of the most critical, and at the same time difficult to control, condition for successful PFM experiment is the state of the probe. Conventional metal-coated silicon tips are extremely prone to contamination, metal peel-off, and wear induced by mechanical forces, conductive heating, and electrochemical reactions at the tip-surface junction. This often results in deterioration of contrast and spatial resolution. The hard conductive coating such as doped diamond often have anorder of magnitude lower conductivities compared to metallic counterparts and are also prone to degradation. Correspondingly, development of wear-resistant whole-metal probes or silicon probes with implanted metal inclusions can greatly increase the reproducibility and quantitativeness of PFM measurements.

A second complication in PFM arises from the electrostatic contribution to the signal. Due to the parabolic bias dependence of electrostatic force, as compared to linear piezoelectric coupling, the latter cannot be separated unambiguously using standard harmonic detection approach. Electrostatic contribution can be minimized using stiff probes for strongly piezoelectric materials. However, imaging weakly piezoelectric and soft materials necessitates the use of cantilevers with small spring constants, for which distributed electrostatic forces dominate the PFM signal. For low frequencies, this distributed electrostatic force can be minimized by using specific locations corresponding to the node of the buckling oscillation of the cantilever;[51] however, this approach is inapplicable for high frequencies and does not minimize local tip-surface electrostatic forces. This problem can be addressed by introducing shielded probes, currently developed in the context of electrochemical SPM[115] and high-resolution Kelvin Probe Imaging.[116]

## VI.2. High Resolution PFM

To date, PFM has been reported with extremely high resolution of at least below 10 nm level. While superior to most property-sensitive SPM techniques, there is no fundamental limitation on achieving sub-nm and potentially atomic resolution. As illustrated by the single chemical bond example in Section II, the expected level of response is well within the detection limit of modern SPM systems. The primary difficulty in achieving this goal is the minimization of the electrostatic response to the measured signal and precise control of tip-surface contact area required to achieve molecular and atomic resolution.

## VI.3. Dynamic and High Frequency Phenomena in Materials

In general, the materials response to external stimuli is described by complex stiffnesses. While the real part of the indentation modulus is related to elastic behavior, imaginary part is related to the losses in material, whether it is domain wall motion, viscoelasticity, etc. Local probing of these parameters can provide valuable information on the materials properties; however, SPM always probes the convolution of the probe dynamics with materials dynamics. Deconvolution of this data using improved data acquisition and probe design is one of the prominent tasks for most SPMs.

## VI.4. Quantitative Vector PFM Imaging

As discussed above, electromechanical response of the surface is generally a vector having three independent components. The beam-deflection system used in most commercial SPMs allows only normal and lateral components of the response to be detected necessitating imaging with sample rotation; also, longitudinal displacement of the surface can contribute to the vertical signal. In addition, the sensitivities of vertical and lateral PFM are generally different, necessitating complex and time consuming calibration. This limitation is a fundamental characteristic of cantilever-based force sensor and the use of alternative configurations for force sensor, e.g. recently introduced 3D SPM[78] or 3-axis nanoindentors[117] can provide a future alternative.

## VI.5. Theory

One of the limitations of PFM is the lack of quantitative theoretical tools for the description of image formation mechanism. While the cantilever dynamics can be well described using elastic beam model and the relevant theory is well developed in the context of various SPM techniques, the description of vertical and lateral voltage-dependent contact mechanics of piezoelectric materials represents an extremely complex problem. Currently, rigorous solutions are available only for a very limited number of materials symmetries. The high relevance of this field to the broad range of inorganic and biological systems will undoubtedly stimulate further development of theoretical approaches based both on simplified analytical models and numerical Green's function methods.

## VII. PFM - Beyond the Electromechanics

In this review, we have discussed in detail electromechanical imaging of piezoelectric and ferroelectric materials, domain switching and spectroscopy based ultimately on bulk models for material behavior. However, with the increase of spatial resolution, SPM becomes more and more sensitive to surface properties of material. This tendency is enhanced by the dielectric constant mismatch between bulk ferroelectric and the surface layer, resulting in disproportionate contribution of surface layer to materials properties. At the same time, ferroelectric surfaces are a treasure throve of novel physical and chemical phenomena related to the presence of switchable polarization. Ferroelectric electron emission has been well known for several decades; recently an approach for cold fusion was suggested using similar phenomena.[118] Variable temperature measurements of surface potential above ferroelectric surfaces have established the fact that polarization is screened by mobile charges in air[119] and allowed kinetic and thermodynamics parameters of screening process to be established.[120] These observations have been corroborated by detailed surface studies.[121] Chemical reactivity of ferroelectric surfaces in acid dissolution reaction has long been known to be polarization dependent.[122] Recently, similar behavior was observed for metal photodeposition processes.[123] The combination of polarization-dependent chemical reactivity and domain patterning gives rise to ferroelectric lithography – a novel approach for nanoscale structure fabrication.[40,124] Almost completely unexplored is a broad area of surface reactivity during PFM experiment. Shown in Fig. 18 is an example of topographic changes during repetitive PFM imaging on PZT and $BiMnO_3$ surfaces,[125] indicative of electrochemical reaction during imaging. Even broader spectrum of phenomena is observed in the ferroelectric heterostructures, in which tip-induced polarization switching can strongly modify the behavior of second component,[126] giving rise to new materials and devices. The future will undoubtedly see a broad spectrum of novel developments in chemistry, biology, and materials science, enabled by PFM.

## VIII. Acknowledgements

The authors gratefully acknowledge multiple interactions with A. Gruverman, B.J. Rodriguez, A.P. Baddorf, and E.W. Plummer. Research supported by ORNL SEED project (SVK), Oak Ridge National Laboratory, managed by UT-Battelle, LLC, for the U.S. Department of Energy under Contract DE-AC05-00OR22725.

# Figure Captions

**Fig. 1.** Electromechanical phenomena on all length scales in inorganic and biological systems. In inorganic systems, properties measured on the macroscopic scale can be related to the atom structure determined by the diffraction methods on the atomic level. This approach is not applicable for complex biological systems, necessitating local SPM studies of properties from molecular to macroscopic level.

**Fig. 2.** Electromechanical coupling on molecular level. (a) Application of electric field result in bond contraction or expansion. (b) Application of external force results in change of dipole moment. (c) Disordered materials such as poled ceramics, polymers or biological systems can be described as piezoelectric texture.

**Fig. 3.** Angular dependence of vertical (a,c,e) and lateral (b,d,f) piezoresponse (piezoresponse surfaces) for (a,b) $BaTiO_3$, (c,d) $PbTiO_3$ and (e,f) collagen.

**Fig. 4.** Orientation dependence of effective electromechanical response as a function of the angle between the electric field and symmetry axis. Shown are surfaces for textures with uniform angular distributions with $\theta_c$ = 0, 45, 90, and 135°. Note that disorder simplifies the shape of response surface and results in decrease of piezoelectric coupling. Response is zero for completely disordered ($\theta_c$ = 180°) material.

**Fig. 5.** Annual number of publications on SPM on ferroelectric and piezoelectric materials (courtesy of A. Gruverman).

**Fig. 6.** (a) Force-based SPM can be conveniently described using force-distance curve, showing the regimes in which contact (C), non-contact (NC), intermittent contact (IC), and interleave imaging are performed. Also shown are domains of repulsive and attractive tip-surface interactions. (b) Voltage modulation SPMs can be described using force-distance bias surface. In the small signal limit, signal in techniques such as PFM, AFAM, EFM, and KPFM are directly related to the derivative in bias or distance direction.

**Fig. 7.** (a) Equivalent circuit for the PFM contrast including concentrated force acting on the tip, distributed force acting on the cantilever, and components of surface displacement. Tip-surface junction mechanics is represented using vertical and lateral springs. (b) Simplified equivalent circuit. (c) Components of the response vector. (d) In conventional beam-deflection systems, longitudinal surface displacement is detected as vertical PFM signal.

**Fig. 8.** (a) Vertical and (b) lateral PFM on PMN-PT surface illustrating the presence of antiparallel *c* domains. The lateral contrast is due to surface tilt in the vicinity of domain walls. Corresponding (c) amplitude and (d) angle image and expanded view of (e) region I and (f) region II. Note that depending on orientation, domains walls are "bright" (in-plane response in lateral direction) and "dark" (in-plane response in longitudinal direction). This asymmetry is due to the difference in signal transduction between lateral and longitudinal components of surface displacement.

**Fig. 9.** (a) PFM image of the grid pattern and (b) corresponding FFT image. (c) Wavevector dependence of the FFT peak intensity illustrating the minimal feature size. (d) Wavevector dependence of tip-surface transfer function, illustrating resolution. (e) Writing signal for variable mesh-size grid and (f) corresponding PFM image, illustrating real-space approach for determination of minimal feature size. Unlike the FFT method, in this case minimum feature size is not unambiguous.

**Fig. 10.** (a,c,e) Frequency and bias dependent amplitude response diagram and (b,d,f) the regions of dominant electromechanical contribution of the cantilevers with (a,b) $k = 0.1$ N/m, (c,d) $k = 2.4$ N/m and (e,f) $k = 45$ N/m. (a,c,e) Plotted is log(Amplitude). (b,d,f) White corresponds to the regions with a dominant electromechanical contribution, while black corresponds to regions with a dominant non-local electrostatic contribution.

**Fig. 11.** Frequency-bias spectroscopy for non-piezoelectric $SiO_2$ in (a) non-contact and (b) contact mode. (c,d) Response diagrams for two PZT grains with opposite poplarization orientation. Note the different trend in frequency dependence of nulling potential. (e) Piezoresponse x-signal and (f) error map after subtraction of linear component.

**Fig. 12.** (a) Dependence of the vertical and lateral response amplitudes for $SiO_2$ and PZT in contact mode. Shown are the vertical response for $SiO_2$ (solid), PZT (dash-dot), lateral response for $SiO_2$ (dash), and PZT (dash-dot-dot). (b) The deviation of the nulling potential from surface potential is directly related to the electromechanical response of the surface.

Frequency dependence of (c) electromechanical and (d) electrostatic response for PZT with a good tip (solid), PZT with a deteriorated tip (dash), $SiO_2$ in the non-contact mode (dash dot), and contact mode (dash dot dot).

**Fig. 13.** Diagram of the experimental set-up for 2D spectroscopic cantilever dynamics measurements. A normal or shear piezo actuator is placed between the stage and the probe tip. (a) The piezo actuator is driven with the ac components of the driving signal and the tip with the dc component. (b) The actuator is grounded and both ac and dc bias is applied to the tip.

**Fig. 14.** (a) Amplitude and (b) phase response of 2D spectroscopic measurement for a cantilever with a spring constant of 4.5 N/m.

**Fig. 15.** Amplitude of the response of 2D spectroscopy on cantilevers with spring constants of (a) 40 N/m and (b) 0.15 N/m.

**Fig. 16.** (a) Experimental set-up for piezoelectric nanoindentation. (b) Typical time dependence of applied load and (c) load-displacement and (d) piezoresponse-displacement curves for polycrystalline PZT sample.

**Fig. 17.** Piezoelectric nanoindentation on polycrystalline $BaTiO_3$. (a) Load-displacement and (b) piezoresponse-displacement curves. (c,d) PNI mapping of $BaTiO_3$. 20 X 20 indents with 2 micron step was performed. Measured harmonic displacement was averaged for load (c) 0 – 0.5 mN and (d) during hold at 3 mN.

**Fig. 18.** (a,c) Surface topography and (b,d) PFM image of BiMnO$_3$ surface before (a,b) and after (c,d) repetitive PFM writing. Note that long-term bias exposure results in local electrochemical reaction as evidenced by formation of topographic feature and decay of piezoresponse amplitude. (e) surface topography and (f) PFM images of PZT surface exhibiting similar behavior.

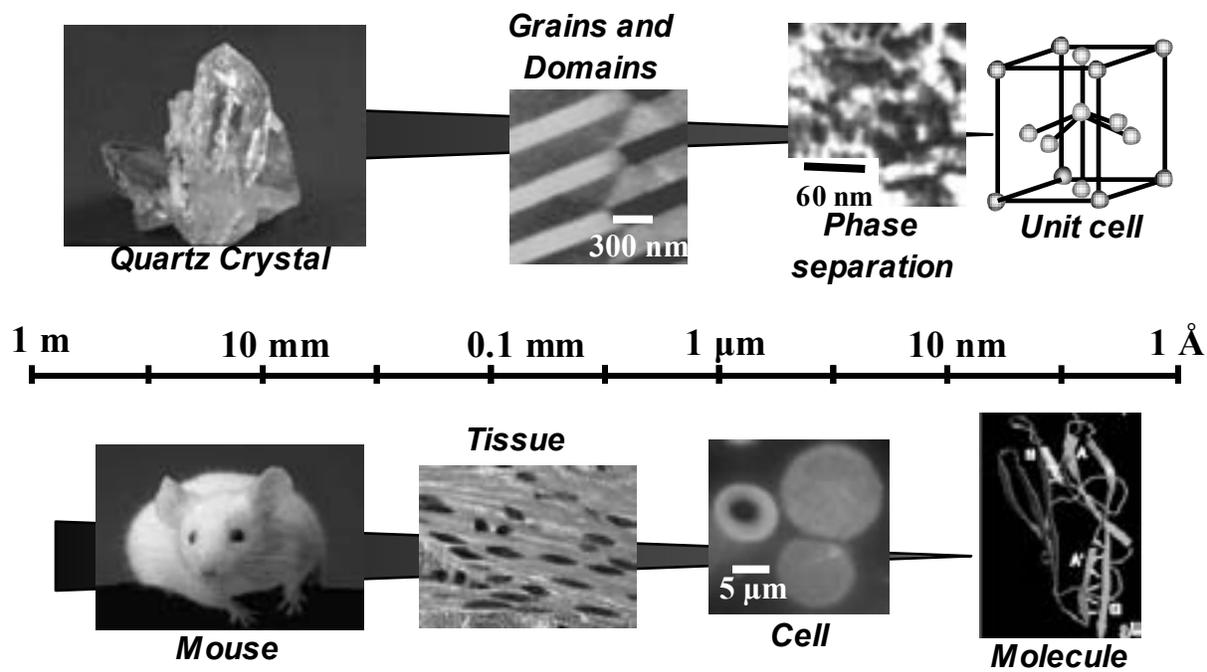

**Fig. 1.** S.V. Kalinin, A. Rar, and S. Jesse

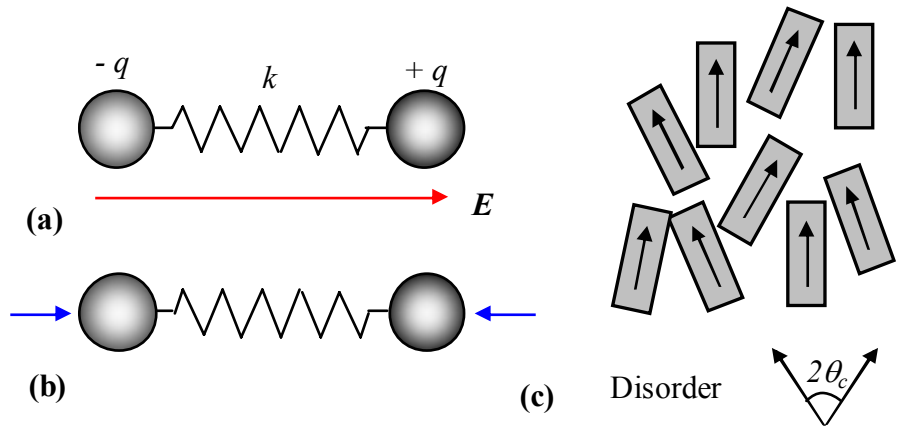

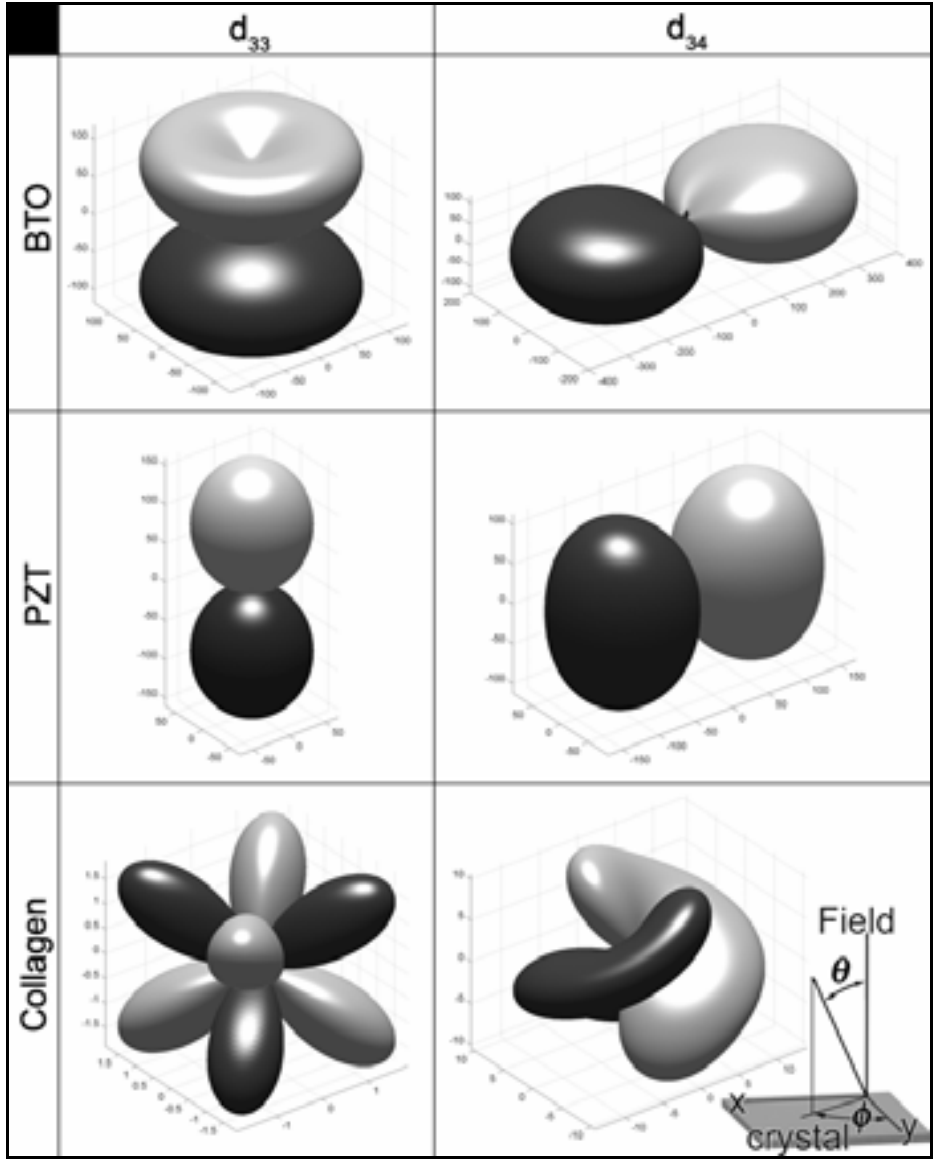

**Fig. 3.** S.V. Kalinin, A. Rar, and S. Jesse

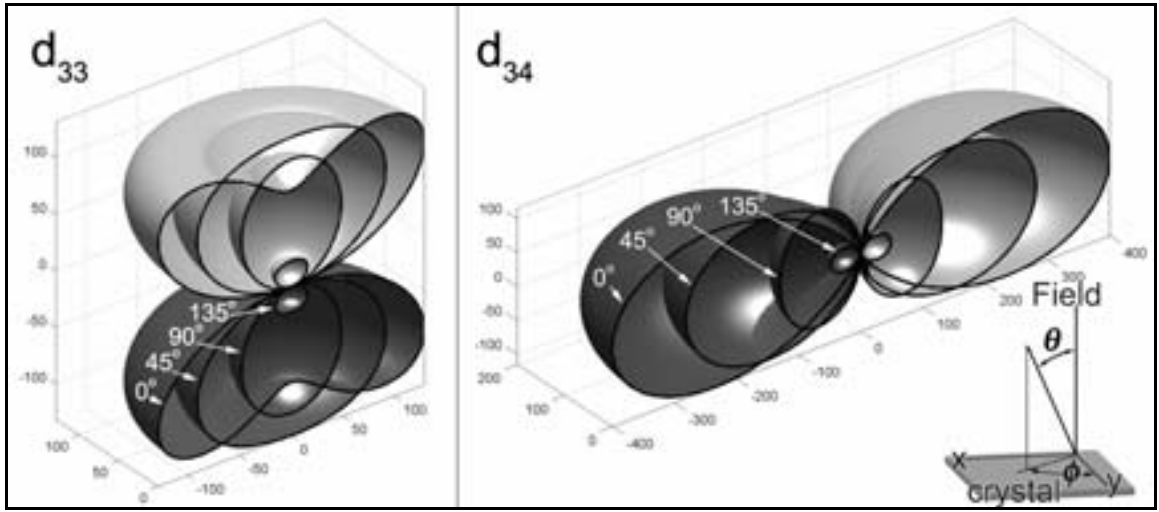

**Fig. 4.** S.V. Kalinin, A. Rar, and S. Jesse

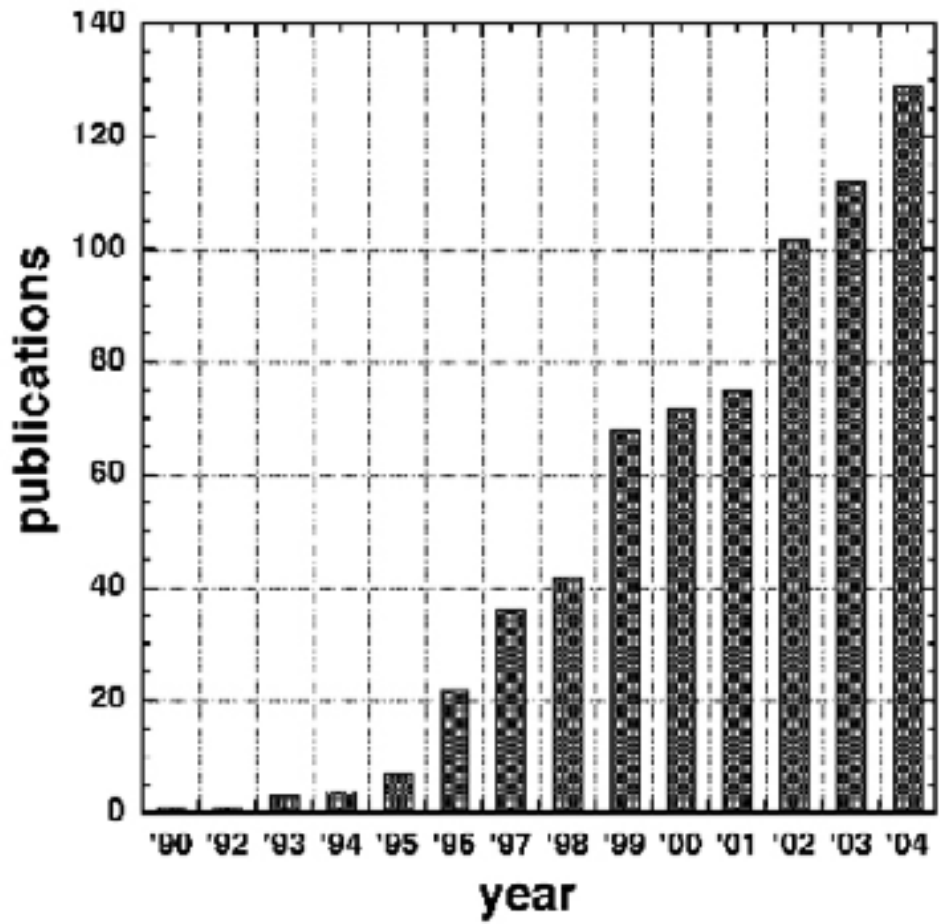

**Fig. 5.** S.V. Kalinin, A. Rar, and S. Jesse

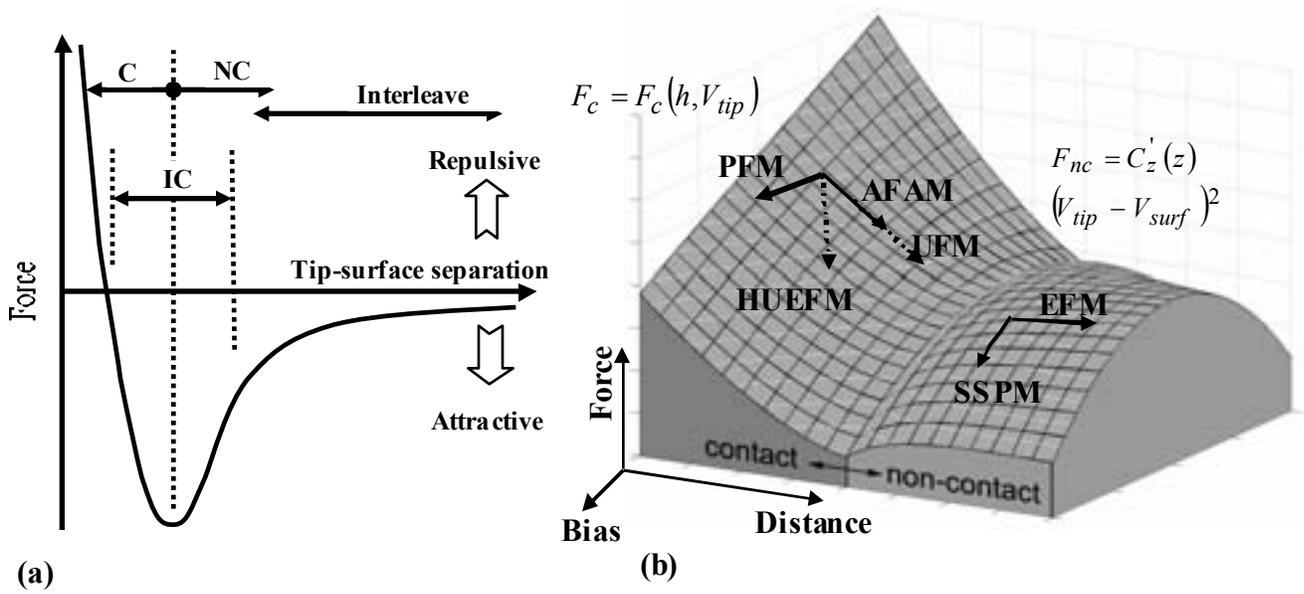

**(a)**

**(b)**

$F_c = F_c(h, V_{tip})$

$F_{nc} = C_z^{'}(z)(V_{tip} - V_{surf})^2$

**Fig. 6.** S.V. Kalinin, A. Rar, and S. Jesse

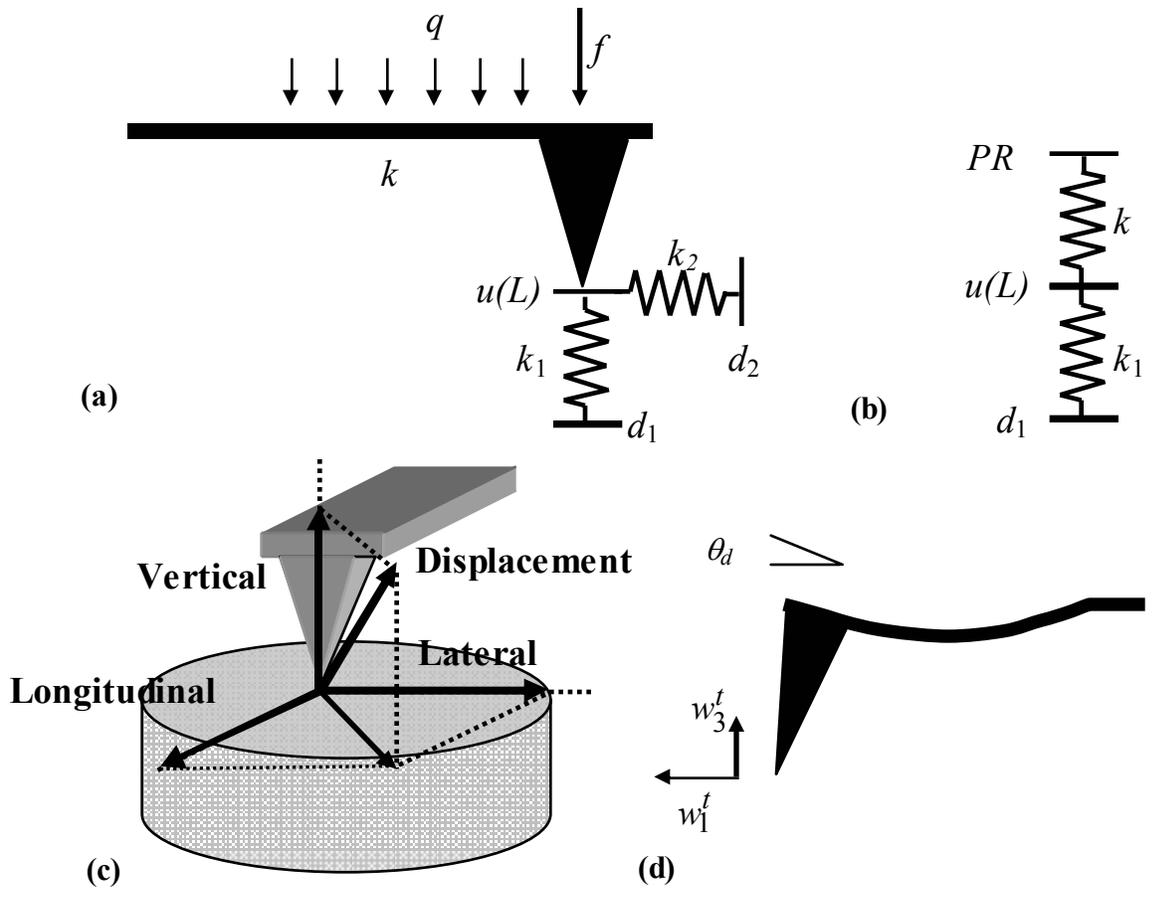

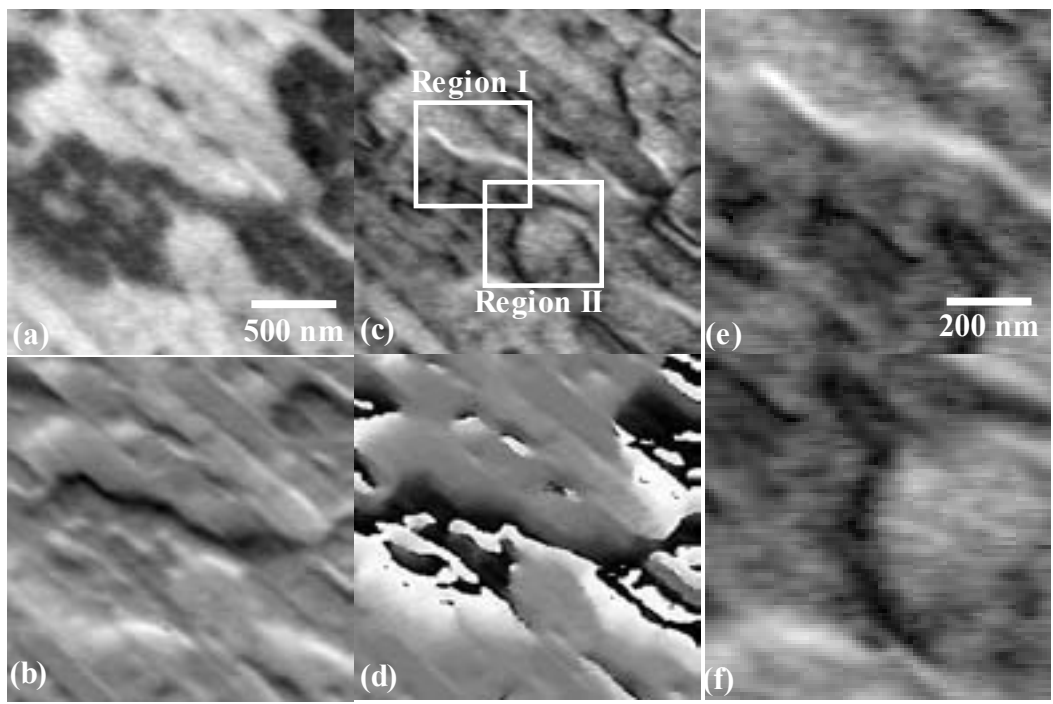

Region I

Region II

(a)    500 nm    (c)    (e)    200 nm

(b)    (d)    (f)

**Fig. 8.** S.V. Kalinin, A. Rar, and S. Jesse

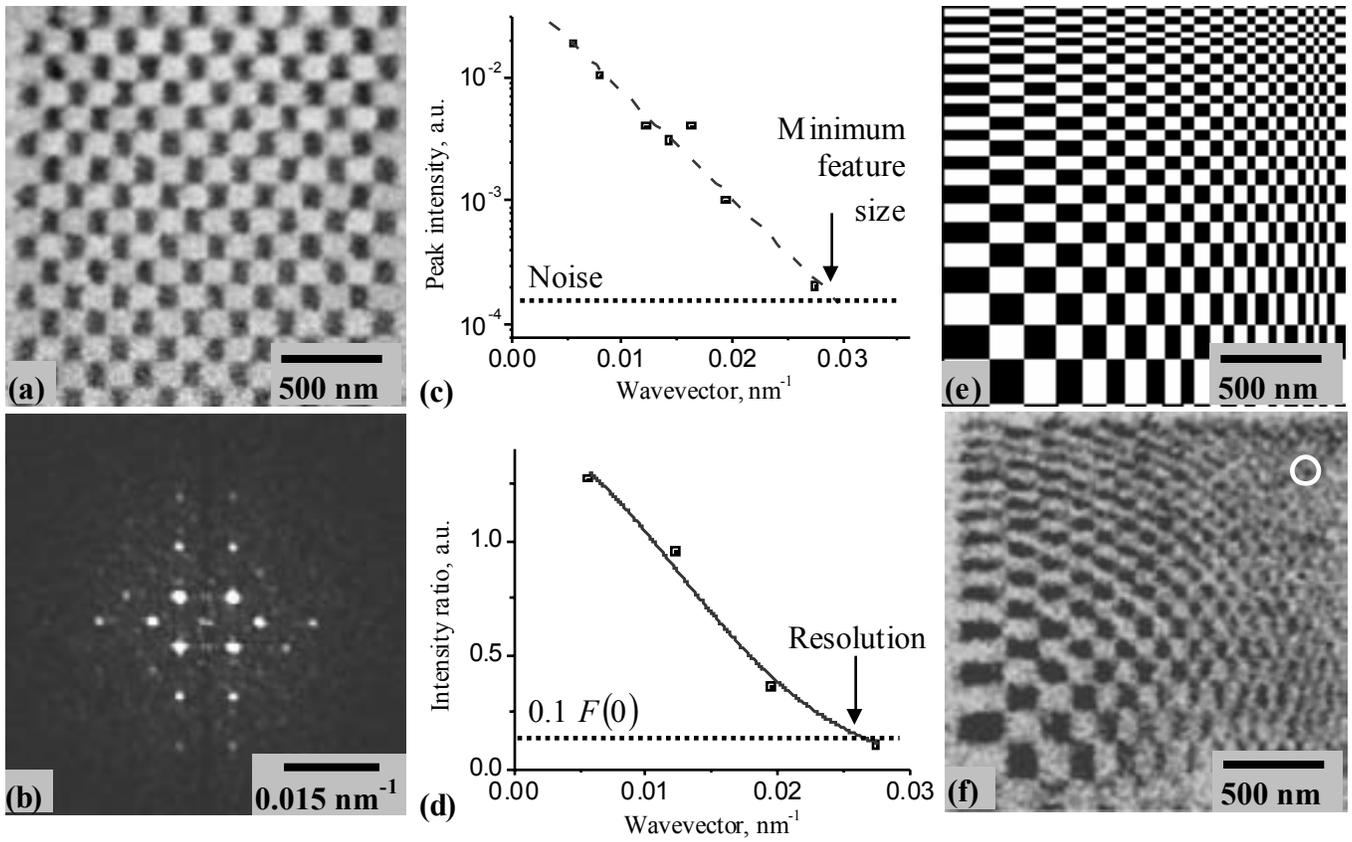

**Fig. 9.** S.V. Kalinin, A. Rar, and S. Jesse

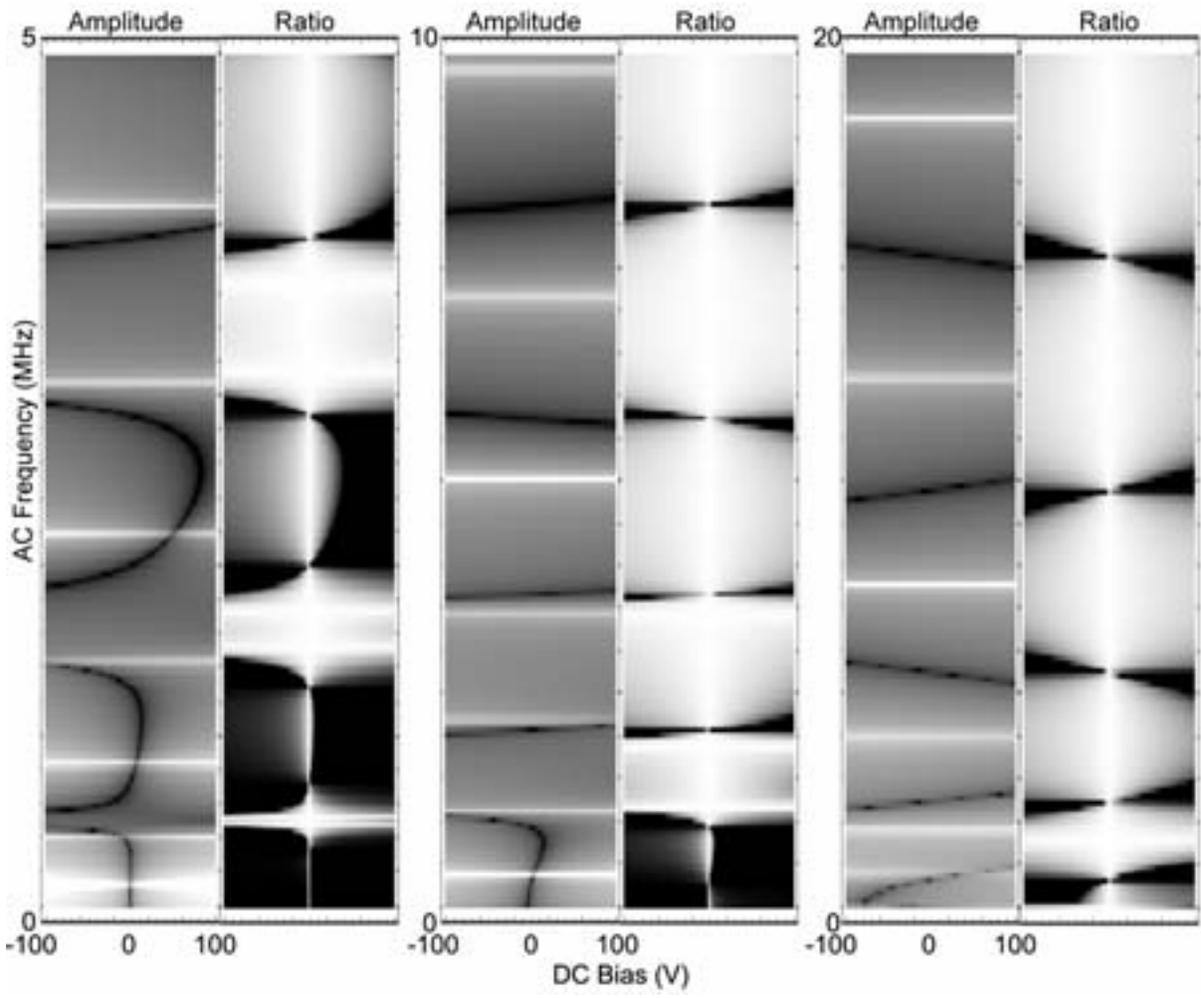

**Fig. 10.** S.V. Kalinin, A. Rar, and S. Jesse

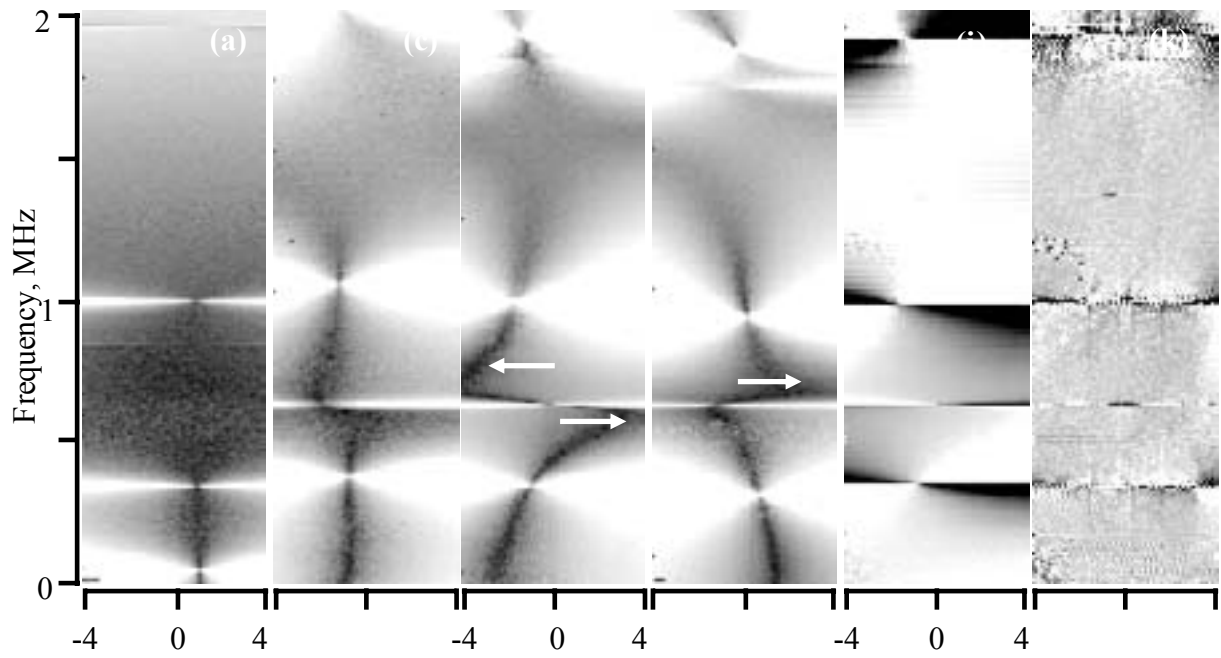

**Fig. 11.** S.V. Kalinin, A. Rar, and S. Jesse

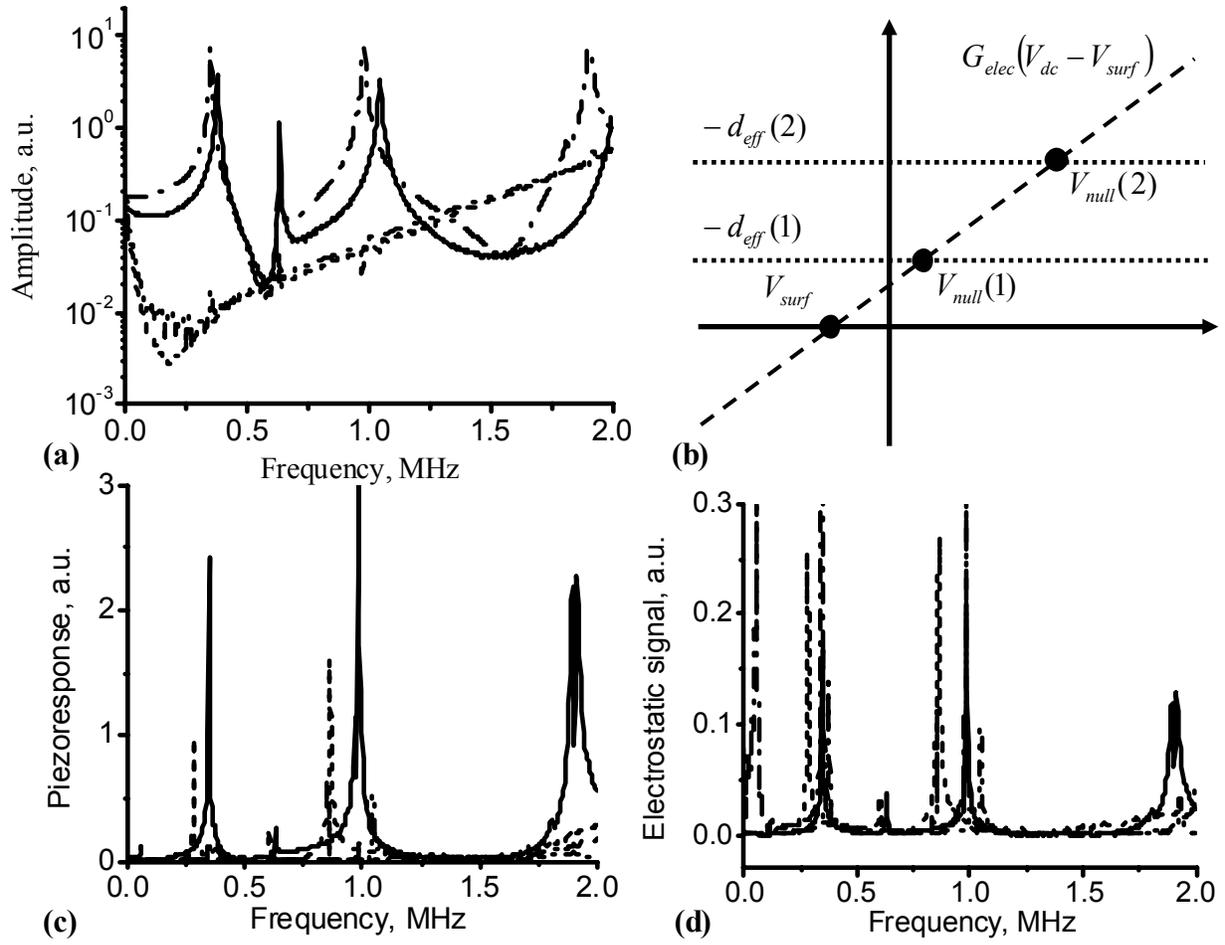

**(a)**

**(b)** $G_{elec}(V_{dc} - V_{surf})$

$-d_{eff}(2)$

$-d_{eff}(1)$

$V_{surf}$

$V_{null}(1)$

$V_{null}(2)$

**(c)**

**(d)**

**Fig. 12.** S.V. Kalinin, A. Rar, and S. Jesse

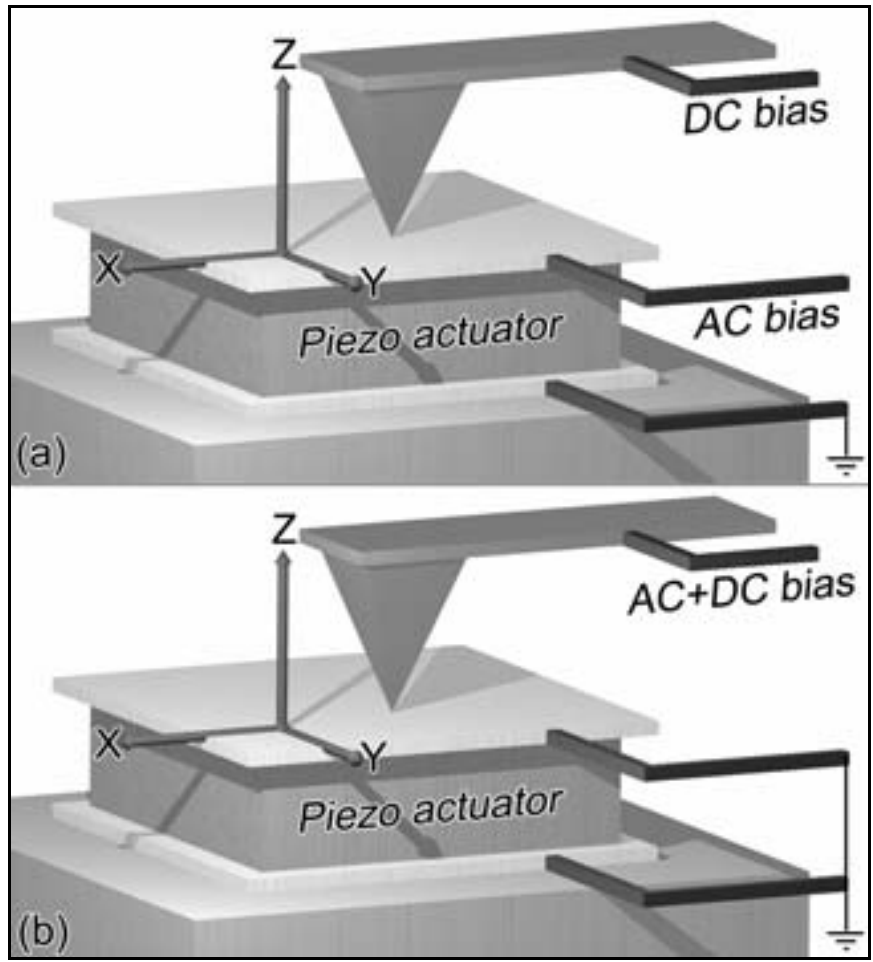

**Fig. 13.** S.V. Kalinin, A. Rar, and S. Jesse

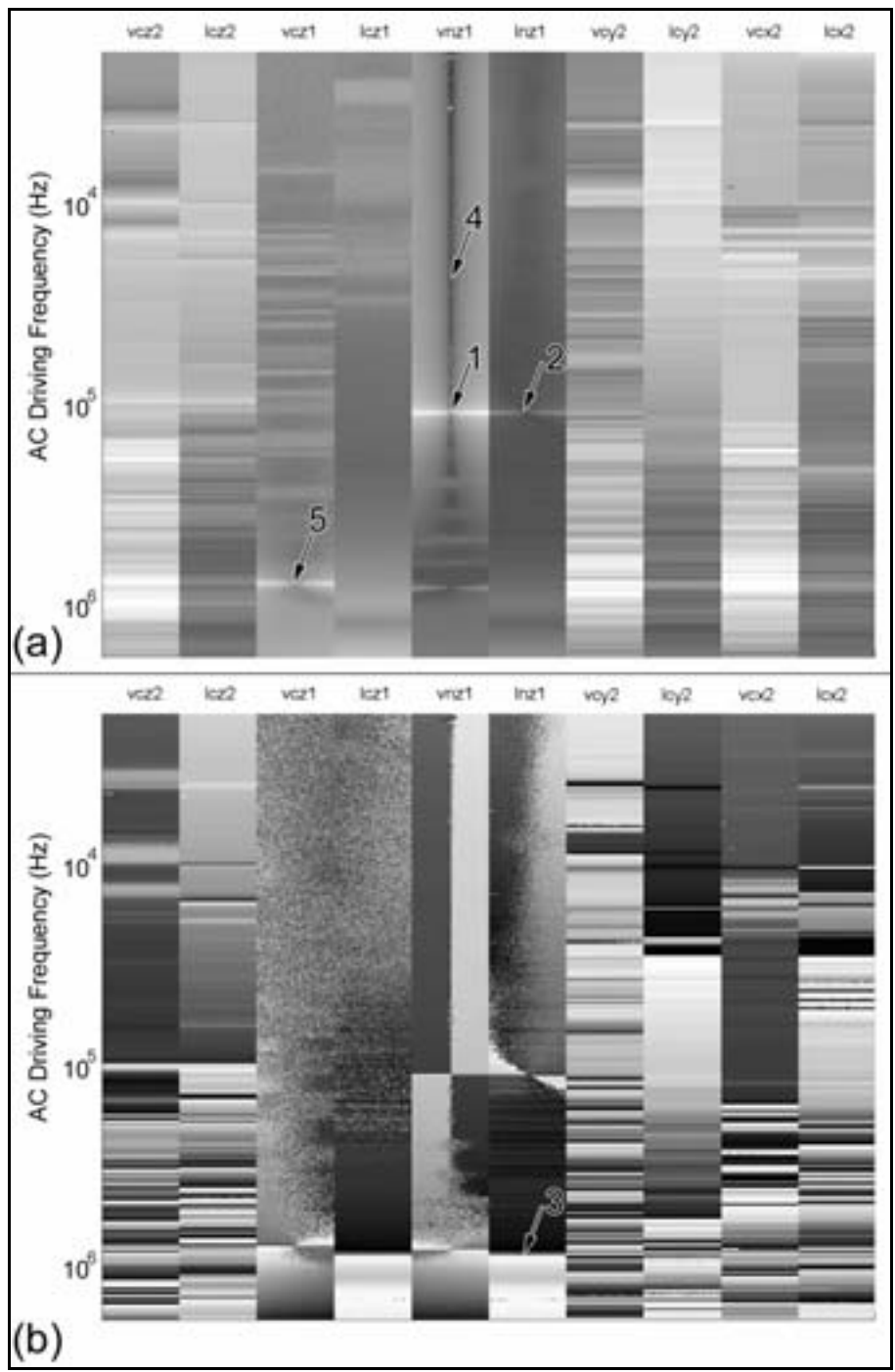

**Fig. 14.** S.V. Kalinin, A. Rar, and S. Jesse

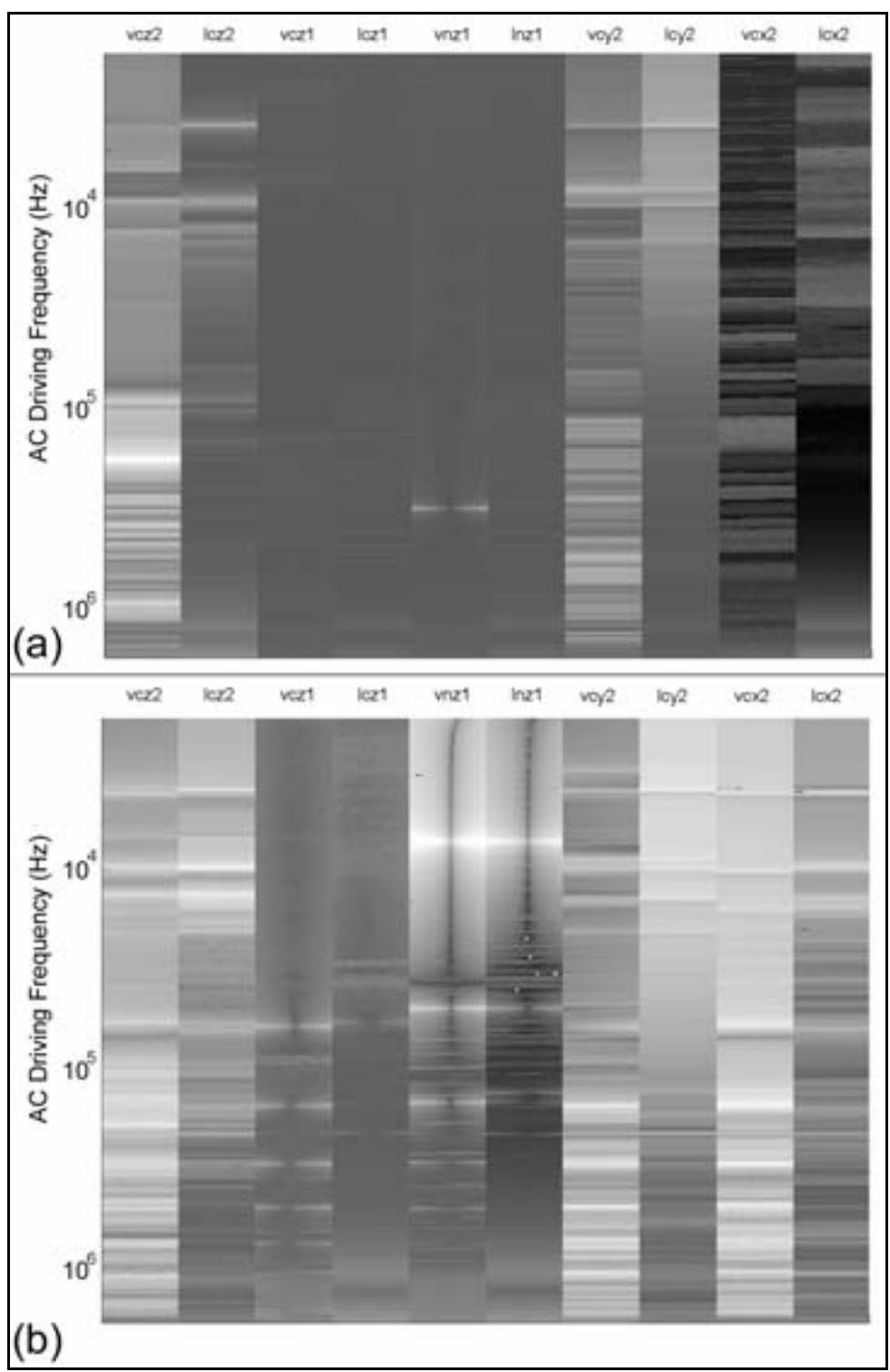

**Fig. 15.** S.V. Kalinin, A. Rar, and S. Jesse

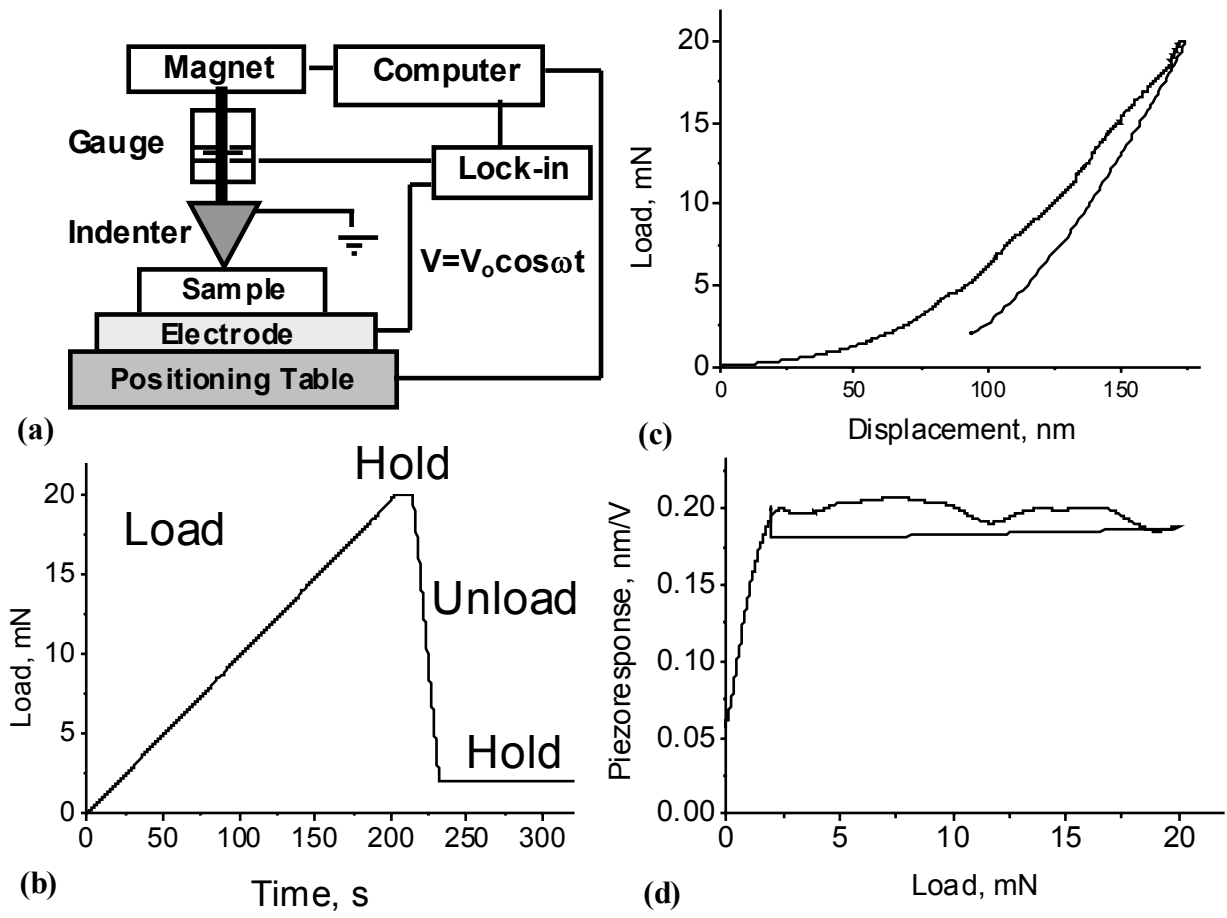

**Fig. 16.** S.V. Kalinin, A. Rar, and S. Jesse

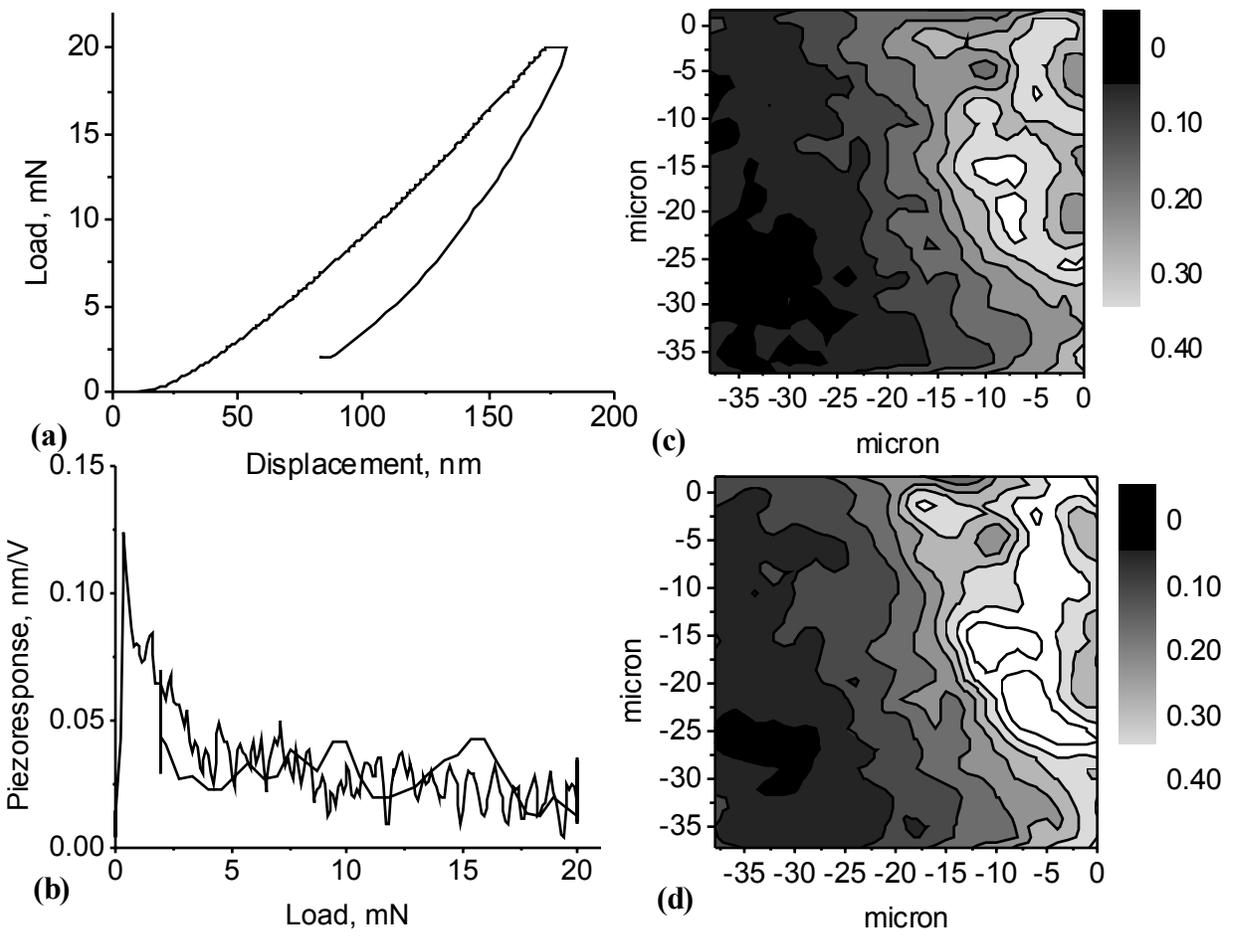

**Fig. 17.** S.V. Kalinin, A. Rar, and S. Jesse

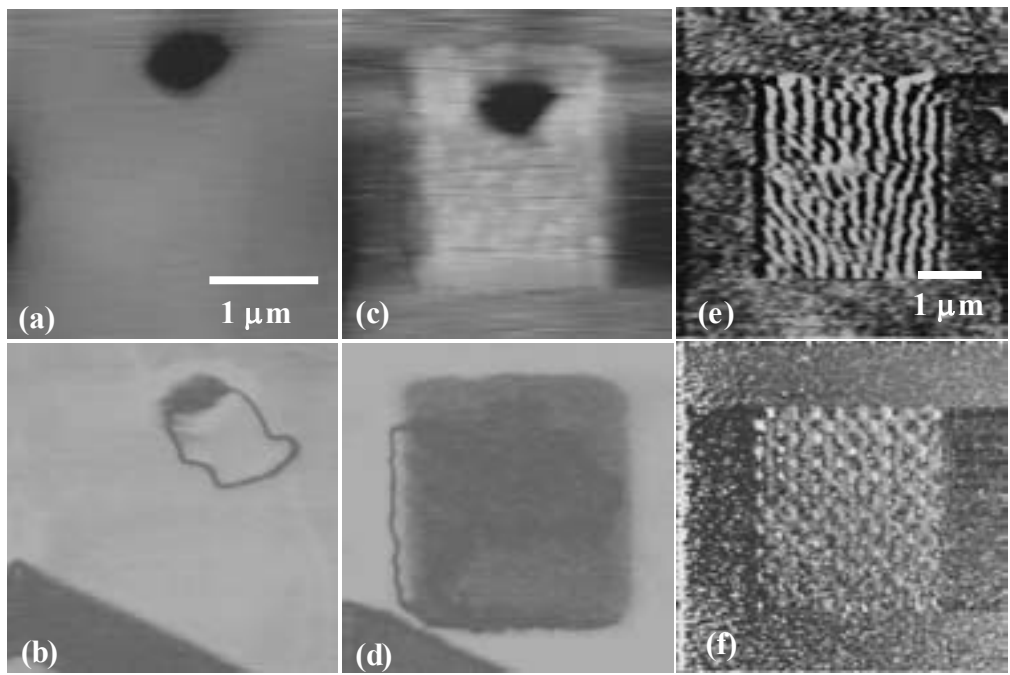

**Fig. 18.** S.V. Kalinin, A. Rar, and S. Jesse

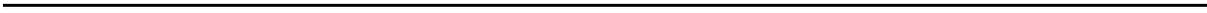